\documentclass[12pt]{article}
\usepackage{amsthm, amssymb, amsmath}
\usepackage{graphicx}
\usepackage{natbib}
\usepackage{multirow,enumerate, url}
\usepackage{float,threeparttable}

\newcommand{\blind}{0}
\newtheorem{remark}{Remark}
\newtheorem{thm}{Theorem}
\newtheorem{lem}{Lemma}

\newtheorem{defi}{Definition}[section]

\usepackage[ruled]{algorithm2e}
\usepackage{algorithmic}

\usepackage[bookmarks=true,
	colorlinks,%
	linkcolor=blue,%
	citecolor=blue,%
	plainpages=false,%
	pdfstartview=FitH
	]{hyperref}
\bibpunct{(}{)}{;}{a}{}{,}

\def\rE{{\mathrm E}}

\def\cF{{\cal F}}

\def\hG{{\hat G}}

\def\Gn{{\mathbb G}_n}

\def\bI{{\boldsymbol I}}

\def\hJ{{\hat J}}

\def\rP{\boldsymbol{\mathrm P}}

\def\Pn{\Bbb{P}_n}

\def\bR{\mathbb{R}}

\def\tT{\tilde{T}}
\def\bu{{\boldsymbol u}}

\def\bU{{\boldsymbol U}}

\def\bv{{\boldsymbol v}}
\def\bbv{{\bar \bv}}
\def\bV{{\boldsymbol V}}
\def\bbV{{\bar \bV}}

\def\tV{{\tilde{V}}}

\def\bW{{\boldsymbol W}}
\def\tbW{{\widetilde \bW}}

\def\bx{{\boldsymbol{x}}}

\def\bX{{\boldsymbol X}}

\def\by{{\boldsymbol y}}

\def\bZ{{\boldsymbol Z}}

\def\eps{\epsilon}

\def\argmin{\mathop{\rm argmin}}
\def\real{\mathop{{\rm I}\kern-.2em\hbox{\rm R}}\nolimits}

\def\trans{^{\rm T}}

\def\trace{\mbox{trace}}

\def\balpha{\boldsymbol \alpha}

\def\hbalpha{\hat{\boldsymbol \alpha}}
\def\btheta{\boldsymbol \theta}
\def\bbeta{\boldsymbol \beta}
\def\hbbeta{\hat{\boldsymbol \beta}}

\def\bata{\boldsymbol \eta}

\def\hbata{\hat{\boldsymbol \eta}}
\def\bgamma{\boldsymbol{\gamma}}

\def\hbgamma{\hat{\bgamma}}

\def\baromega{\bar \omega}

\def\hbtheta{\hat{\boldsymbol \theta}}

\def\bxi{\boldsymbol \xi}

\def\bzeta{\boldsymbol{\zeta}}
\def\hbzeta{\hat{\bzeta}}

\def\sgn{\hbox{sgn}}
\def\Var{\hbox{Var}}

\def\bone{{\boldsymbol 1}}
\def\bzero{{\boldsymbol 0}}
\def\type1{\mbox{Type-\uppercase\expandafter{\romannumeral1}}}

\newcommand{\E}{\textrm{E}}

\newcommand{\lkonv}{\stackrel{\cal{L}}{\longrightarrow}}

\def\pvalue{\mbox {p-value}}

\begin{document}

\def\spacingset#1{\renewcommand{\baselinestretch}%
{#1}\small\normalsize} \spacingset{1}

\date{}
\if0\blind
{
  \title{\bf Robust subgroup-classifier learning and testing in change-plane regressions}
  \author{Xu Liu$^1$,
    Jian Huang$^2$, Yong Zhou$^3$ and Xiao Zhang$^{4,}$\thanks{
    	CONTACT zhangxiao1994@cuhk.edu.cn; The Chinese University of Hong Kong-Shenzhen, China}
\hspace{.2cm}\\
    $^1$Shanghai University of Finance and Economics, Shanghai, China\\
    $^2$The Hong Kong Polytechnic University, Hong Kong SAR, China\\
    $^3$East China Normal University, Shanghai, China\\
    $^4$The Chinese University of Hong Kong-Shenzhen,
    Shenzhen, China
    }
  \maketitle
} \fi

\if1\blind
{
  \bigskip
  \bigskip
  \bigskip
  \begin{center}
    {\LARGE\bf Robust subgroup-classifier learning and testing in change-plane regressions}
\end{center}
  \medskip
} \fi

\bigskip
\begin{abstract}
Considered here are robust subgroup-classifier learning and testing in change-plane regressions with heavy-tailed errors, which can identify subgroups as a basis for making optimal recommendations for individualized treatment. A new subgroup classifier is proposed by smoothing the indicator function, which is learned by minimizing the smoothed Huber loss. Nonasymptotic properties and the Bahadur representation of estimators are established, in which the proposed estimators of the grouping difference parameter and baseline parameter achieve sub-Gaussian tails. The hypothesis test considered here belongs to the class of test problems for which some parameters are not identifiable under the null hypothesis. The classic supremum of the squared score test statistic may lose power in practice when the dimension of the grouping parameter is large, so to overcome this drawback and make full use of the data's heavy-tailed error distribution, a robust weighted average of the squared score test statistic is proposed, which achieves a closed form when an appropriate weight is chosen. Asymptotic distributions of the proposed robust test statistic are derived under the null and alternative hypotheses. The proposed robust subgroup classifier and test statistic perform well on finite samples, and their performances are shown further by applying them to a medical dataset. The proposed procedure leads to the immediate application of recommending optimal individualized treatments.
\end{abstract}

\noindent%
{\it Keywords:}  Gaussian-type deviation; 
Heavy-tailed data;
Nonstandard test;
Robust classifier;
Subgroup classifier;
Subgroup detection.
\vfill

\newpage
\spacingset{1.9} 

\section{Introduction}\label{sec:intro}
When studying the risk of a disease outcome, there could be heterogeneity across subgroups characterized by covariates, meaning that the same treatment in different subpopulations may cause different treatment effects of predictors. In the presence of population heterogeneity in classical models, learning the subgroup classifier and testing the existence of subgroups associated with the risk-model heterogeneity are important for understanding better the different effects of predictors and modeling better the association of diseases with predictors. In precision medicine, this plays a core role in guiding personalized treatment to individuals in a population by identifying subgroups with different treatment effects on disease. There has been much previous research on learning the subgroup classifiers of individuals based on various models 
\cite{2011Subgroup, 2018The, 2020Threshold, 2021Multithreshold, 2021Single}.

Before learning the subgroup classifier, it is necessary to test for the existence of subgroups of individuals to address the potential risk of finding false-positive subgroups. This necessity not only arises from the data themselves but is also intrinsic to statistics, because the nonexistence of subgroups causes the identifiability problem when learning the subgroup classifier. 
However, this test problem belongs to the class of nonstandard tests with loss of identifiability under the null hypothesis; see \cite{1943Tests,1994Optimal, 1995Admissibility,1977Hypothesis, 2009On, Liu2022a,kang2024inference}, among others. Therefore, the focus herein is on learning the subgroup classifier and testing for the existence of subgroups simultaneously, which offers more information and the potential for making the best recommendations for optimal individualized treatments and guiding future treatment modification and development.

Let $\{\bV_i=(y_i, \bX_i,\bZ_i,\bU_i), i=1,\cdots, n\}$ be the observed data, which are $n$ independent and identically distributed copies of $\bV=(y, \bX,\bZ,\bU)$. Consider the regression model with change plane \cite{2011Testing,2021Single,Mukherjee2022,Liu2022a}
\begin{align}\label{model}
	y_i=\bX_i\trans\balpha+\bZ_i\trans\bbeta\bone(\bU_i\trans\bgamma\geq0) + \eps_i,
\end{align}
where $\balpha=(\alpha_1,\cdots,\alpha_p)\trans\in\Theta_{\alpha}\subseteq\mathbb{R}^{p}$, $\bbeta=(\beta_1,\cdots,\beta_q)\trans\in\Theta_{\beta}\subseteq\mathbb{R}^{q}$ and $\bgamma=(\gamma_1,\cdots,\gamma_r)\trans\in\Theta_{\gamma}\subseteq\mathbb{R}^r$ are unknown parameters, and $\rE(\eps_i|\bX_i,\bZ_i,\bU_i)=0$ and $\rE(|\eps_i|^{2+\delta}|\bX_i,\bZ_i,\bU_i)=M_{\delta}<\infty$ for some $\delta\geq0$. When $\delta=0$, the error has a finite second moment, denoted by $M_0=\rE(\eps_i^2|\bX_i,\bZ_i,\bU_i)$. For easy expression, let $\btheta=(\balpha\trans,\bbeta\trans)\trans$. Following the expressions in \citet{Liu2022a}, $\bU$ is called the {\it grouping variable}, $\bgamma$ is called the {\it grouping parameter}, $\bZ$ is called the {\it grouping difference variable}, $\bbeta$ is called the {\it grouping difference parameter}, $\bX$ is called the {\it baseline variable}, and $\balpha$ is called the {\it baseline parameter}. Herein, the indicator function $\bone(\bU\trans\bgamma\geq0)$ is called the {\it subgroup classifier}.

The technology for collecting and processing data sets has improved considerably in recent years, and one is now more likely to encounter heavy-tailed or low-quality data, thereby causing the typical assumption of a Gaussian or sub-Gaussian distribution to fail. Therefore, new challenges arise compared with the classic methodology for modeling non-Gaussian or heavy-tailed data. Even for linear regression models with heavy-tailed errors, the ordinary least squares (OLS) estimators are suboptimal both theoretically and empirically. Instead, proposed herein is a robust estimator of subgroup classification by considering the change-plane model \eqref{model} with heavy-tailed errors. This paper addresses two important problems for model \eqref{model} with heavy-tailed errors, i.e., subgroup-classifier learning (\autoref{sec:learning}) and subgroup testing for whether subgroups exist (\autoref{sec:test}).

\subsection{Robust subgroup-classifier learning}\label{sec:intro_learning}

\citet{2021Multithreshold} considered the change-plane model \eqref{model} with Gaussian errors. Also, \citet{2021Single} investigated a quantile regression with a change plane and derived the asymptotic normalities for the grouping difference parameter and the grouping parameter. However, although quantile or median regression models require no Gaussian or sub-Gaussian assumption, they essentially estimate the conditional quantile or median regression instead of the conditional mean regression. If the mean regression is of interest in practice, then these procedures are not feasible unless the error distribution is symmetric around zero, which may be too strong to cause the misspecification problem. See \cite{Fan2017} for some examples that demonstrate the distinction between conditional mean regression and conditional quantile or median regression.

Linear regression models with heavy-tailed errors are prevalent in the literature. \citet{Fan2017} proposed a robust estimator of high-dimensional mean regression in the absence of asymmetry and with light tail assumptions. \citet{Zhou2018} provided a robust M-estimation procedure with applications to dependence-adjusted multiple testing. \citet{Sun2020} and \citet{Wang2021} studied adaptive Huber regression for linear regression models with heavy-tailed errors. \citet{Chen2020} investigated robust inference via multiplier bootstrap in multiple response regression models, constructing robust bootstrap confidence sets and addressing large-scale simultaneous hypothesis testing problems.

\begin{figure}[!h]
	\begin{center}
		\includegraphics[height=2.7in,width=1.9in, angle=-90]{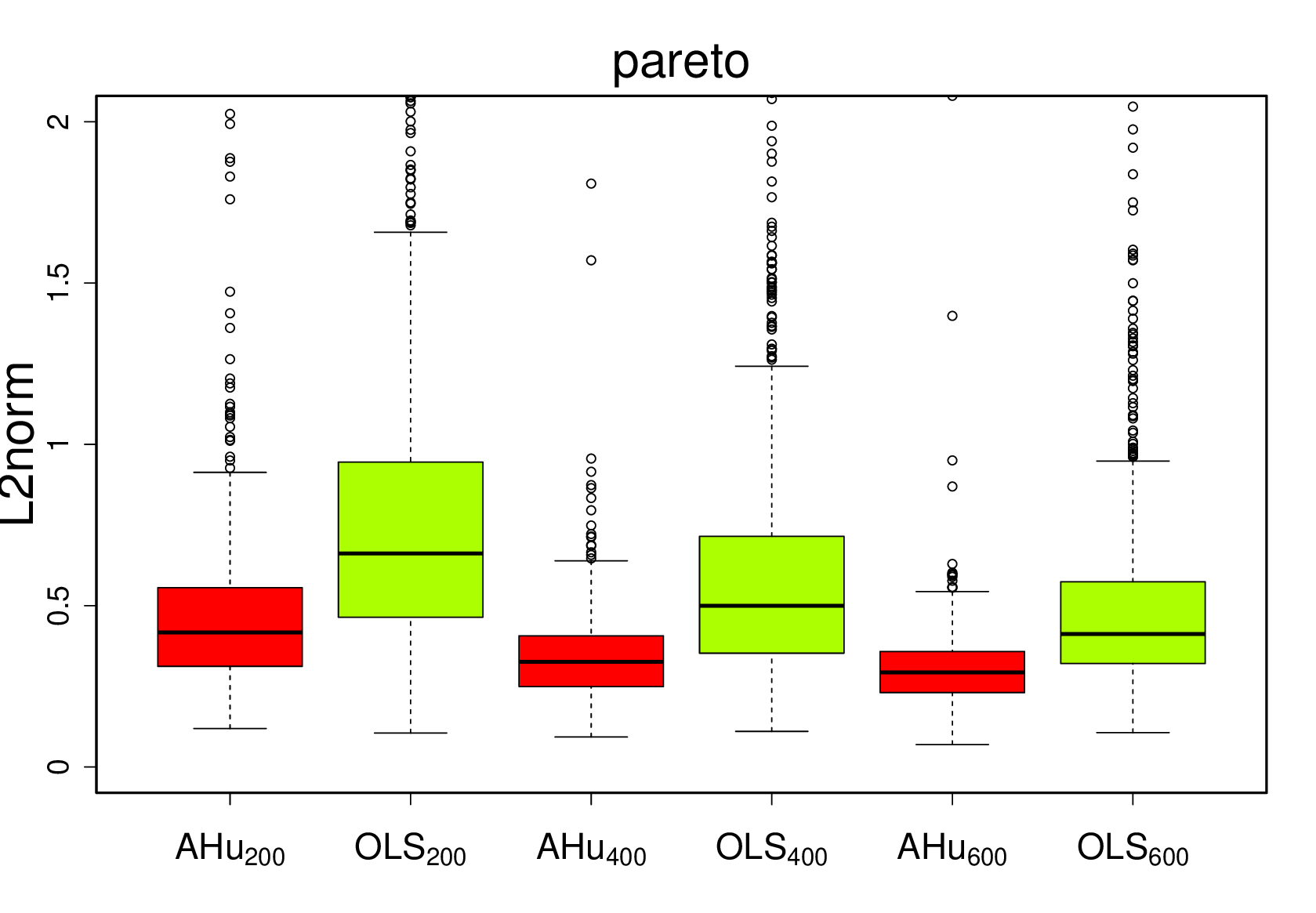}
		\includegraphics[height=2.7in,width=1.9in, angle=-90]{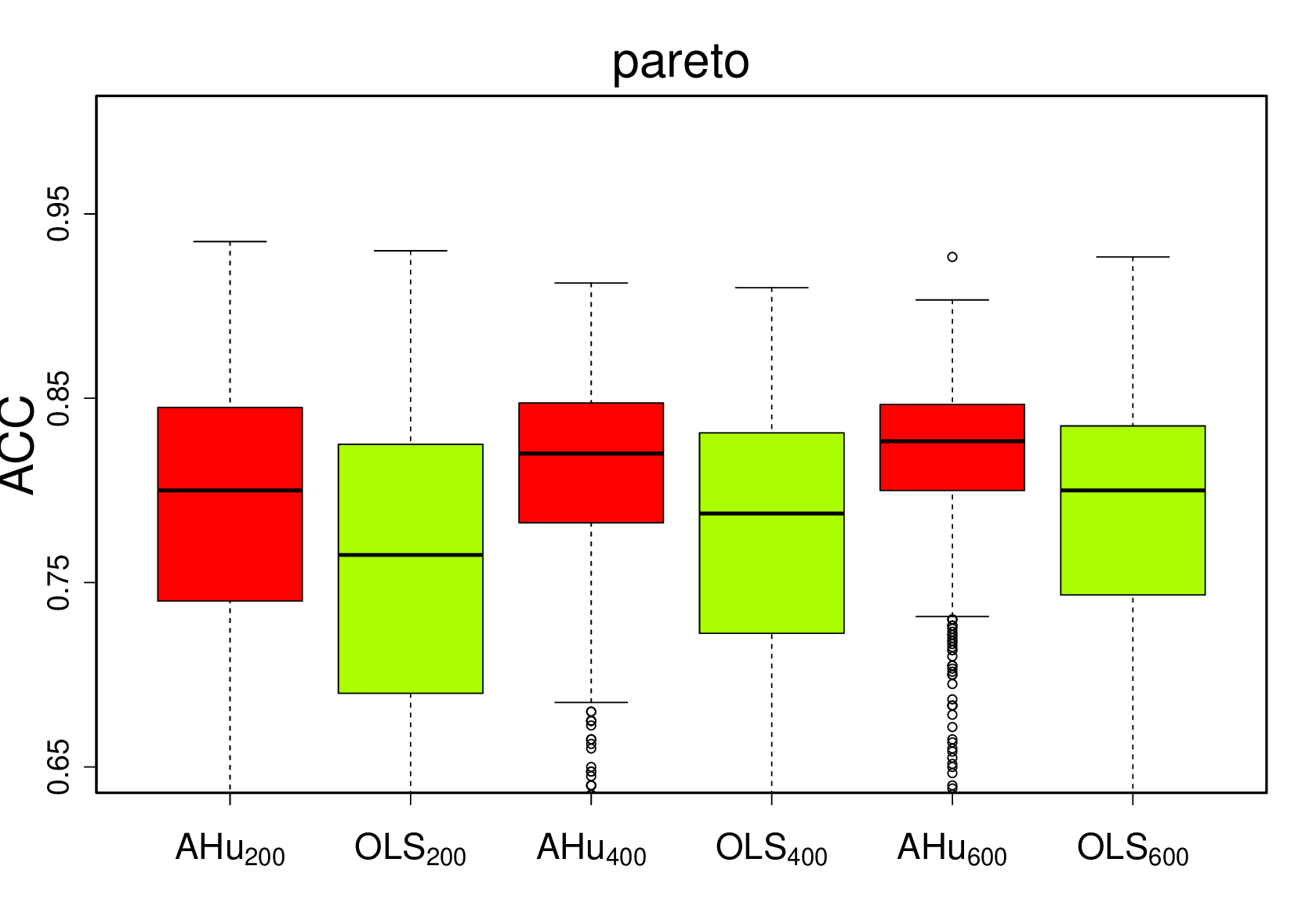}
		\caption{\it Estimation errors of parameter $\bbeta$ with $L_2$ norm (left) and accuracies of estimated subgroup classifier (right) for heavy-tailed errors generated from the Pareto distribution.}
		\label{fig_intro_est}
	\end{center}
\end{figure}

\citet{Mukherjee2022} studied the change-plane problem under heavy-tailed errors when $\balpha=0$ and $\bX=1$, which is a special case of the change-plane model \eqref{model}, and they left the general change-plane model with heavy-tailed errors for future work. Herein, a new robust procedure is introduced to estimate parameters and consequently to learn the subgroup classifier. \autoref{fig_intro_est} shows boxplots of the estimation errors of the parameter $\btheta=(\balpha\trans, \bbeta\trans)\trans$ with $L_2$ norm and the accuracies of the estimated subgroup classifier, where the $L_2$ norm of the estimation errors is defined as $\|\hbtheta_{\tau,h}-\btheta^*\|$ with the robust estimator $\hbtheta_{\tau,h}$ of the true grouping parameter $\btheta^*$, and the accuracy is defined as $\mbox{ACC}=1-n^{-1}\sum_{i=1}^{n}\left|\bone(\bU_{i}\trans\hbgamma_{\tau,h}\geq0)-\bone(\bU_{i}\trans\bgamma^*\geq0)\right|$ with the robust estimator $\hbgamma_{\tau,h}$ of the true parameter $\bgamma^*$. Here, the settings are $(p, q, r)=(3, 3, 3)$ and $n=(200, 400,600)$ with 1000 repetitions,
the heavy-tailed errors are generated from the Pareto distribution $Par(2,1)$ with shape parameter 2 and scale parameter 1, and $X_1=Z_1=1$ and $(X_{2},\cdots,X_{p})\trans=(Z_{2},\cdots,Z_{p})\trans$ and $(U_{2},\cdots,U_{r})\trans$ are generated independently from multivariate normal distributions $N(\bzero_{p-1},\sqrt{3}\bI_{p-1})$ and $N(\bzero_{r-1},\sqrt{3}\bI_{r-1})$, respectively; see \autoref{sec:sim} for details. As used by \citet{2021Single}, the smooth function $K(u)=\{1+\exp(-u)\}^{-1}$ with smoothness parameter $h=\sqrt{\log (n)/n}$ is chosen, and the proposed robust estimation procedure (AHu) is compared with the method based on OLS \citep{2021Multithreshold}. \autoref{fig_intro_est} sends the important message that in the presence of heavy tails, compared with the existing method \citep{2021Multithreshold}, the proposed robust estimators not only reduce the estimation error dramatically but also improve significantly the accuracy of the subgroup classifier.

\subsection{Robust subgroup testing}\label{sec:intro_test}

Another goal of this paper is to test for the existence of subgroups, i.e.,
\begin{align}\label{test}
	H_0: \bbeta=\bzero\quad versus\quad H_1: \bbeta\neq \bzero.
\end{align}
Note that the grouping parameter $\bgamma$ is not identifiable under the null hypothesis.

The classic Wald-type test or score-based test is powerful in standard test problems when there is no identifiability problem in both the null and alternative hypotheses, but these common procedures are not feasible when nuisance parameters are present. \citet{1994Optimal} and \citet{1995Admissibility} studied the weighted average exponential form, which was originally introduced by \citet{1943Tests}. \citet{1977Hypothesis} investigated well the supremum of the squared score test (SST) statistic for mixture models, which was applied by \citet{2009On} and \citet{2017Subgroup} to semiparametric models in censoring data.

All the aforementioned testing methods are optimal tests based on the weighted average power criterion. However, because these optimal tests take the weighted exponential average of the classical tests over the grouping parametric space $\Theta_{\gamma}$, they may not only not perform well in practice when the dimension of $\Theta_{\gamma}$ is large but also give rise to a heavy-burden calculation of the $\pvalue$ or the critical value. Instead, \citet{Liu2022a} introduced a new test statistic by taking the weighted average of the SST (WAST) over $\Theta_{\gamma}$ and removing both the inverse of the covariance and the cross-interaction terms to overcome the drawbacks of SST. Thanks to its closed form, WAST achieves more-accurate type-I errors and significantly improved power and hence dramatically reduced computational time as a byproduct. 

\begin{figure}[!h]
	\begin{center}
		\includegraphics[scale=0.35]{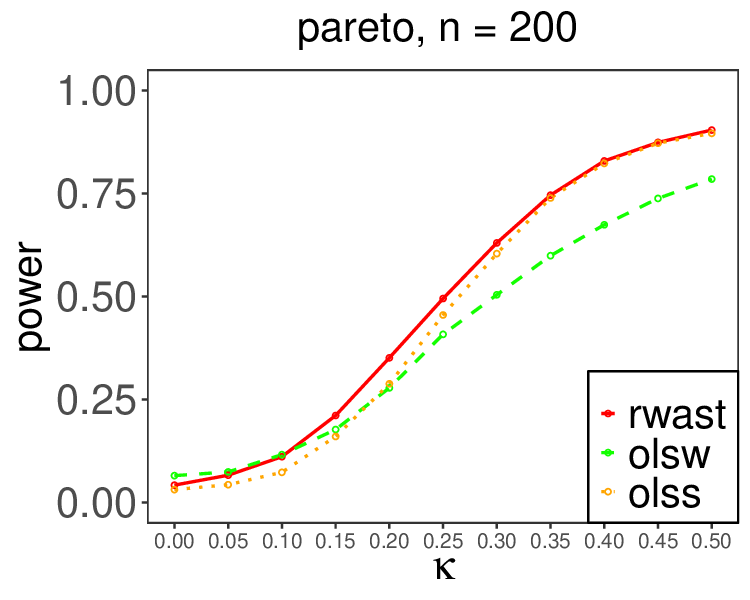}
		\includegraphics[scale=0.35]{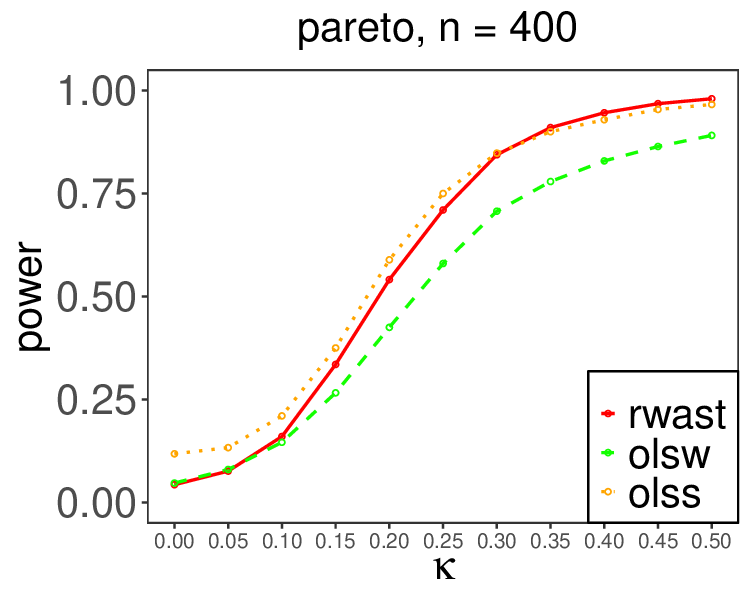}
		\includegraphics[scale=0.35]{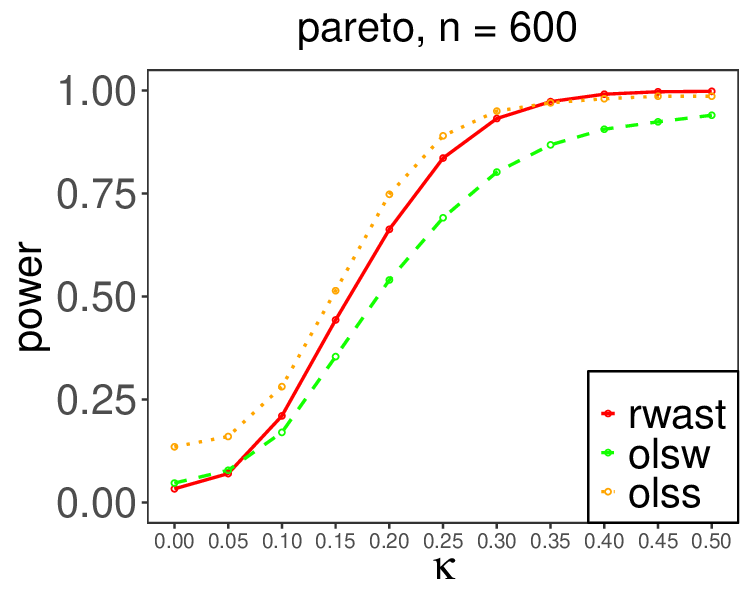}
		\caption{\it Powers of test statistics by proposed RWAST (rwast, red solid line), WAST (olsw, green dashed line), and SST (olss, blue dotted line) for $(p, q, r)=(3, 3, 3)$ and generating heavy-tailed errors from the Pareto distribution. 
		}
		\label{fig_intro_test}
	\end{center}
\end{figure}

All the aforementioned test procedures require the important assumption of Gaussian or sub-Gaussian errors in the change-plane models, none of which apply to heavy-tailed data sets. Therefore, proposed herein is a robust test procedure based on WAST \citep{Liu2022a}, called robust WAST (RWAST). \autoref{fig_intro_test} shows the power curves of the proposed RWAST, WAST as introduced by \citet{Liu2022a}, and SST as considered by \citet{1977Hypothesis, 2017Subgroup}. A total of 1000 bootstrap samples is set, and the other settings are the same as those in \autoref{sec:intro_learning}. With the nominal significance level $\alpha=0.05$, the type-I errors for these three methods are
$(\mbox{rwast},\mbox{olsw},\mbox{olss})=(0.042, 0.065, 0.031)$ for $n=200$, $(\mbox{rwast},\mbox{olsw},\mbox{olss})=(0.043, 0.047, 0.118)$ for $n=400$, and $(\mbox{rwast},\mbox{olsw},\mbox{olss})=(0.033, 0.047, 0.135)$ for $n=600$. It follows that RWAST controls the type-I errors well, while those of SST based on quadratic loss are larger and deviate far from 0.05. \autoref{fig_intro_test} shows that in the presence of heavy tails, the proposed RWAST achieves larger power in comparison with WAST based on the ordinary quadratic loss.

In summary, from the demonstration examples in \autoref{sec:intro_learning} and \autoref{sec:intro_test}, compared with the existing nonrobust methods for heavy-tailed data, the proposed robust estimation procedure is characterized by lower estimation errors and higher accuracy, and the robust test procedure has more-accurate type-I errors and larger power.

\subsection{Main contributions}

The main contributions of this paper are summarized as follows. First, the robust estimation procedure for the change-plane model \eqref{model} with heavy-tailed errors is investigated carefully. The proposed robust estimator adapts to the sample size, the robustification parameter in the Huber loss, the smoothness parameter when approximating the indicator function in the subgroup classifier, and the moments of errors. The sacrifices made in pursuit of robustness and smoothness are analyzed theoretically, with the bias involving the robustification parameter arising from the pursuit of robustness, and the one involving the smoothness parameter arising from the approximation to the indicator function. The nonasymptotic properties for parameters $\balpha$ and $\bbeta$ are established, as well as those for the grouping parameter. The theoretical results reveal that the proposed estimators of the grouping difference parameter and baseline parameter have Gaussian-type deviations \citep{Devroye2016}. Also provided is the nonasymptotic Bahadur representation of the proposed robust estimators, which is convenient for deriving the classical asymptotic results needed for statistical inference such as hypothesis tests and constructing confidence regions. 
Extensive simulation studies show that the proposed robust estimation procedure is superior to its several competitors.

Second, for the change-plane model with heavy-tailed errors, RWAST is proposed, which makes full use of the data's heavy-tailed information and overcomes the drawbacks of loss of power in practice and the heavy computational burden of SST when the dimension of the grouping parameter is large. The asymptotic distributions of the proposed RWAST under the null and alternative hypotheses are established based on the theory of degenerate U-statistics. As with exponential average tests, the proposed asymptotic distributions are not standard (e.g., the normal or $\chi^2$ distribution), so a novel bootstrap method that is easily implemented and theoretically guaranteed is introduced to mimic the critical value or $\pvalue$. Comprehensive simulation studies conducted with finite sample sizes and for various heavy-tailed error distributions show the excellent performance of the proposed RWAST, which improves the power significantly and reduces the computational burden dramatically.

In summary, a novel robust estimator is proposed that adapts to the sample size, dimension, robustification parameter, moments, and smoothness parameter in pursuit of the optimal tradeoff among bias, robustness, and smoothness. To the best of the author's knowledge about change-plane analysis, the literature contains no nonasymptotic results with sub-Gaussian tails for parameters $\btheta$ and $\bata$, and no nonasymptotic results with sub-exponential tails for the Bahadur representation of these parameters. Furthermore, a robust test procedure is proposed that improves on WAST in the change-plane model with heavy-tailed errors.

\subsection{Notation and Organization of paper}

Here, some useful notation is introduced for convenience of expression. For a vector $\bv\in \bR^{d}$ and a square matrix $A=(a_{ij})\in\mathbb{R}^{d\times d}$, denote by $\|\bv\|$ the Euclidean norm of $\bv$, by $\trace(A)=\sum_{i=1}^{d}a_{ii}$ the trace of $A$; $\|\bv\|_{A}^2=\sum_{i, j}a_{ij}v_iv_j$ and $\bv^{\otimes2}=\bv\bv\trans$. Denote by $\|A\|_p=\sup\{\|A\bx\|_p: \bx\in\bR^{d}, \|\bx\|_p=1\}$ the induced operator norm for a matrix $A=(a_{ij})\in\bR^{m\times d}$.

Denote by $\rP$ the ordinary probability measure such that $\rP f=\int f d\rP$ for any measurable function $f$, by $\Pn$ the empirical measure of a sample of random elements from $\rP$ such that $\Pn f=n^{-1}\sum_{i=1}^{n}f(\bV_i)$, and by $\Gn$ the empirical process indexed by a class $\cF$ of measurable functions such that $\Gn f=\sqrt{n}(\Pn-\rP)f$ for any $f\in \cF$. Let $L^p(Q)$ be the space of all measurable functions $f$ such that $\|f\|_{Q, p}:=(Q|f|^p)^{1/p}<\infty$, where $p\in[1, \infty)$ and $(Q|f|^p)^{1/p}$ denotes the essential supremum when $p=\infty$. Let $N(\eps, \cF, \|\cdot\|_{Q, 2})$ be an $\eps$-covering number of $\cF$ with respect to the $L^2(Q)$ seminorm $\|\cdot\|_{Q, 2}$, where $\cF$ is a class of measure functions and $Q$ is finite discrete.


The remainder of this paper is organized as follows. \autoref{sec:learning} provides the robust estimators for the grouping difference parameter and grouping parameter as well as the subgroup classifier, and theorems reveal that these estimators achieve Gaussian-type deviations. Also derived is the Bahadur representation of the robust estimators, and it is shown that the remainder of the Bahadur representation achieves sub-Gaussian tails. \autoref{sec:test} presents the RWAST statistic and establishes its limiting distributions under the null and alternative hypotheses. \autoref{sec:sim} reports the results of simulation studies conducted to evaluate the finite-sample performance of the proposed methods with competitors in the change-plane models with heavy-tailed errors. The performance of the proposed methods is illustrated further by applying them to a medical dataset in \autoref{sec:case}. Finally, \autoref{sec:conclus} concludes with remarks and further extensions. The proofs are provided in the Supplementary Material, and an R package named ``wasthub'' is available at \href{https://github.com/xliusufe/wasthub}{https://github.com/xliusufe/wasthub}.

\section{Robust subgroup-classifier learning}\label{sec:learning}

In this section, the subgroup classifier is learned to partition subjects into two subgroups, and nonasymptotic properties are provided for the robust estimators, whose deviations achieve sub-Gaussian tails. To adapt for different magnitudes of errors and to robustify the estimation, the Huber loss \citep{Huber1964,Fan2017,Wang2021,Han2021} is considered, the definition of which begins this section.

\begin{defi}
	The Huber loss $L_{\tau}(u)$ \citep{Huber1964} is defined as
	\begin{align}\label{eq:loss_huber}
		L_{\tau}(u)
		=\left\{\begin{array}{ll}
			u^2/2 & \text{ if } |u|\leq \tau, \\
			\tau|u|-\tau^2/2& \text{ otherwise,}
		\end{array} \right.
	\end{align}
	where $\tau>0$ is a tuning parameter called the {\it robustification parameter} \citep{Sun2020,Chen2020}, which regulates the bias and robustness.
\end{defi}

The Huber loss is a hybrid of the squared loss with small errors and absolute loss for large errors. Denoting by $L(u)=u^2/2$ the ordinary quadratic loss, it is straightforward to see that $L(u)=\lim_{\tau\rightarrow\infty}L_{\tau}(u)$.

\subsection{Robust estimation}

Rewrite model \eqref{model} as
\begin{align}\label{model1}
	y_i=\bX_i\trans\balpha+\bZ_i\trans\bbeta\bone(U_{1i}+\bU_{2i}\trans\bata\geq0) + \eps_i,
\end{align}
where $\bU_i=(U_{1i},\bU_{2i}\trans)\trans$, $\bata=\gamma_1^{-1}\bgamma_{-1}$ with $\bgamma_{-1}=(\gamma_2,\cdots,\gamma_r)\trans$. To avoid the identifiability problem for $\bata$, $\bbeta\neq\bzero$ is assumed in this section. Denote $\bzeta=(\balpha\trans,\bbeta\trans,\bata\trans)\trans\in\Theta_{\zeta}$, where $\Theta_{\zeta}=\Theta_{\alpha}\times\Theta_{\beta}\times\Theta_{\eta}$ is the product space.

Because the indicator function $\bone(U_1+\bU_2\trans\bata\geq0)$ is not differentiable, it is natural to approximate it by a smooth function $K(u)$ satisfying
\begin{align*}
	\lim_{u\rightarrow+\infty}K(u)=1~\mbox{and}~ \lim_{u\rightarrow-\infty}K(u)=0.
\end{align*}
Note that this smooth function characterizes the cumulative distribution function instead of the density function; see \cite{2007A, 2021Multithreshold, 2021Single, Mukherjee2020} for more details. The literature contains many commonly used smooth functions, such as the cumulative distribution function of standard normal distribution $K(u)=\Phi(u)$, the sigmoid function $K(u) = \{1+\exp(-u)\}^{-1}$, and the mixture of the cumulative distribution function and density of standard normal distribution $K(u)=\Phi(u)+u\phi(u)$. Thus, model \eqref{model1} can be approximated by
\begin{align}\label{model1_smooth}
	y_i=\bX_i\trans\balpha+\bZ_i\trans\bbeta K_h(U_{1i}+\bU_{2i}\trans\bata) + \eps_i,
\end{align}
where $K_h(u)=K(u/h)$, and $h$ is a predetermined tuning parameter associated with $n$ satisfying $\lim_{n\rightarrow\infty}h=0$ , called the {\it smoothness parameter}.

For any $\tau>0$, let $\bzeta_{\tau,h}^*$ be the minimizer defined as
\begin{align}\label{eq:smooth_huber}
	\bzeta_{\tau,h}^*=\argmin_{\bzeta\in\Theta_{\zeta}}\rP L_{\tau}(y-\bX\trans\balpha-\bZ\trans\bbeta K_h(U_1+\bU_2\trans\bata)),
\end{align}
which approximates the minimizer
\begin{align}\label{eq:est_huber}
	\bzeta_{\tau}^*=\argmin_{\bzeta\in\Theta_{\zeta}}\rP L_{\tau}(y-\bX\trans\balpha-\bZ\trans\bbeta\bone(U_1+\bU_2\trans\bata\geq0)).
\end{align}
As did \citet{Sun2020}, $\bzeta_{\tau}^*$ is called the {\it Huber coefficient}, which usually distinguishes from the true parameter $\bzeta^*$. Measured by $\|\bzeta_{\tau}^*-\bzeta^*\|$, the Huber error is caused by the robustification for the heavy-tailed errors, while the distance $\|\bzeta_{\tau,h}^*-\bzeta^*\|$ is a consequence of both robustification and smoothness. \autoref{thm2} reveals that $\|\bzeta_{\tau,h}^*-\bzeta^*\|$ is controlled by both $\tau$ and $h$, with $h$ playing the role of the bandwidth in the nonparametric area.

Minimizing the empirical loss in \eqref{eq:smooth_huber} produces the robust estimator of interest, i.e.,
\begin{align}\label{eq:est_huber_em}
	\hbzeta_{\tau,h}=\argmin_{\bzeta\in\Theta_{\zeta}}\Pn L_{\tau}(y-\bX\trans\balpha-\bZ\trans\bbeta K_h(U_1+\bU_2\trans\bata)).
\end{align}
From \eqref{eq:smooth_huber}, \eqref{eq:est_huber}, and \eqref{eq:est_huber_em}, the total estimation error $\|\hbzeta_{\tau,h}-\bzeta^*\|$ can be decomposed into three parts, i.e.,
\begin{align}
	\underbrace{\|\hbzeta_{\tau,h}-\bzeta^*\|}_{\mbox{Total error}}\leq \underbrace{\|\hbzeta_{\tau,h}-\bzeta_{\tau,h}^*\|}_{\mbox{estimation error}}+\underbrace{\|\bzeta_{\tau,h}^*-\bzeta_{\tau}^*\|}_{\mbox{smooth error}}+\underbrace{\|\bzeta_{\tau}^*-\bzeta^*\|}_{\mbox{Huber error}}.
\end{align}

It is natural to use the alternating strategy to obtain the estimate, denoted by $\hbzeta = (\hbalpha\trans, \hbbeta\trans, \hbata\trans)\trans$. Specifically, the parameters $(\balpha\trans, \bbeta\trans)\trans$ and $\bata$ can be estimated iteratively as follows. For given $\bata^{(k)}$, $\balpha^{(k+1)}$ and $\bbeta^{(k+1)}$ are obtained by minimizing
\begin{align*}
	\Pn L_{\tau}(y-\bX\trans\balpha-\bZ\trans\bbeta K_h(U_1+\bU_2\trans\bata^{(k)})),
\end{align*}
and for given $\balpha^{(k+1)}$ and $\bbeta^{(k+1)}$, $\bgamma^{(k+1)}$ is estimated by minimizing
\begin{align*}
	\Pn L_{\tau}(y-\bX\trans\balpha^{(k+1)}-\bZ\trans\bbeta^{(k+1)} K_h(U_1+\bU_2\trans\bata)).
\end{align*}
Iterating these two maximizers leads to the desired robust estimator. The above alternating strategy is summarized in 
Algorithm A in Appendix A of the Supplementary Material.

\subsection{Nonasymptotic properties}\label{properties_est}

This section begins with assumptions needed to establish the nonasymptotic properties. Let $\bbV$ be $\bV$ by removing $U_1$, i.e., $\bbV=(y,\bX,\bZ,\bU_2))$, and $\baromega(\bata)=U_1+\bU_2\trans\bata$ and $\baromega=\baromega(\bata^*)$.
\begin{enumerate}[({A}1)]
	\item The conditional random vectors $\rE(\bX|\bU_2)\in\mathbb{R}^{p}$ and $\rE(\bZ|\bU_2)\in\mathbb{R}^{q}$ given $\bU_2$ are sub-Gaussian, and $\bU_2$ is sub-Gaussian. There is a universal constant $K_1>0$ satisfying $\|\rE(\bX|\bU_2)\|_{\psi_2}\leq K_1$, $\|\rE(\bZ|\bU_2)\|_{\psi_2}\leq K_1$, and $\|\bU_2\|_{\psi_2}\leq K_1$. For any $\bu_2\in\mathbb{R}^{r-1}$, the matrices $\rE(\bX\bX\trans|\bU_2=\bu_2)$, $\rE(\bX\bX\trans|\bU_2=\bu_2)$, and $\rE(\bU_2\bU_2\trans)$ are uniformly positive definite, and there is a universal constant $K_0>0$ satisfying $\lambda_{\min}(\rE(\bX\bX\trans|\bU_2=\bu_2))\geq K_0$, $\lambda_{\min}(\rE(\bZ\bZ\trans|\bU_2=\bu_2))\geq K_0$, and $\lambda_{\min}(\rE(\bU_2\bU_2\trans))\geq K_0$.
	\item The error variable $\eps$ is independent of $(\bX\trans,\bZ\trans,\bU\trans)\trans$ and satisfies $\rE(\eps)=0$ and $\rE(|\eps|^{2+\delta})=M_{\delta}<\infty$ with $\delta\geq0$. Denote $M_0=\rE(|\eps|^{2})$. \item $0<\rE[\bone(\bU\trans\bgamma\geq0)]<1$ for any $\bgamma\in\Theta_{\gamma}$, and there is a constant $\delta_{u_2}>0$ satisfying $\sup_{\bata\in\mathbb{S}^{r-1}}\rE[\bU_2\trans\bata\bone(\bU_2\trans\bata\geq0)]\geq\delta_{u_2}$.
	\item For almost every $\bu_2$, the density of $U_1$ conditional on $\bU_2=\bu_2$ is everywhere positive. The conditional density $f_{\varpi|\bar{\bv}}(\varpi)$ of $\baromega$ given $\bbV$ has continuous derivative, and there is a constant $\kappa_{f}>0$ such that $f_{\varpi|\bar{\bv}}(\varpi)$ and $|f_{\varpi|\bar{\bv}}'(\varpi)|$ are uniformly bounded from above by $\kappa_{f}$ over $(\varpi,\bar{\bv})$. $f_{\varpi|\bar{\bv}}(0)$ is uniformly bounded from below by $\delta_{f_0}$ over $\bar{\bv}$, and $F_{\varpi|\bar{\bv}}(0)$ is uniformly bounded from above by $\kappa_{F}$ over $\bar{\bv}$, where $\delta_{f_0}>0$ and $0<\kappa_F<1$ are constants.
	\item The smooth function $K(\cdot)$ is twice differentiable and $K(-t)=1-K(t)$. $K'(\cdot)$ is symmetric around zero. Moveover, there is a universal constant $\kappa_k>0$ satisfying 
	$\max\left\{\sup_{t\in\mathbb{R}}|K'(t)|,\sup_{t\in\mathbb{R}}|K''(t)|,\int |K'(t)|^jdt, \int |K''(t)|^jdt, \int |t||K'(t)|^jdt\right\}\leq\kappa_k$ with $j=1,2$.
\end{enumerate}

\begin{remark}
	Assumptions~(A1)--(A5) are mild conditions for deriving the nonasymptotic bounds in Theorems~\ref{thm1}--\ref{thm4} below. Assumption~(A1) is the moment condition for covariates. Assumption~(A2) is imposed to control the moment of error and to yield the adaptive nonasymptotic upper bounds; see \cite{Fan2017,Wang2021,Han2021}. Assumption~(A3) is mild and easily verified in practice, and it is usually imposed in change-plane analysis; see \cite{2017Subgroup,Liu2022a}. Assumption~(A4) is required to establish nonasymptotic properties in dealing with the indicator function; see \cite{horowitz1993,2021Single}. By Lemmas~C1--C3 in Appendix C of the Supplementary Material, Assumption~(A5) holds for commonly used smoothing functions such as (i) the cumulative distribution function of standard normal distribution $K(u)=\Phi(u)$, (ii) the sigmoid function $K(u) = \{1+\exp(-u)\}^{-1}$, and (iii) the function $K(u)=\Phi(u)+u\phi(u)$; see \cite{horowitz1993,2021Single}.
\end{remark}

\begin{thm}\label{thm1}
	Let $\btheta=(\balpha\trans,\bbeta\trans)\trans$. If Assumptions~(A1)--(A4) hold, then for some $\delta\geq0$, the minimizer $\bzeta_{\tau}^*=((\btheta_{\tau}^*)\trans,(\bata_{\tau}^*)\trans)\trans$ given in \eqref{eq:est_huber} satisfies
	\begin{align*}
		\|\btheta_{\tau}^*-\btheta^*\|
		\leq5K_1\frac{2M_{\delta}+C\|\btheta^*\|^{2+\delta}K_1^{2+3\delta/2}}{K_0(1-\delta_F)(1+\delta)\tau^{1+\delta}}
	\end{align*}
	and
	\begin{align*}
		\|\bata_{\tau}^*-\bata^*\|
		\leq&\frac{160K_1^2}{K_0\delta_{f_0}\delta_{u_2}\|\bbeta^*\|^2}\left\{\frac{2M_{\delta}+C\|\btheta^*\|^{2+\delta}K_1^{2+3\delta/2}}{K_0(1-\delta_F)(1+\delta)\tau^{1+\delta}}\right\}^2,
	\end{align*}
	where $C>0$ is a constant.
\end{thm}

\autoref{thm1} states that the Huber error is of order $\tau^{-(1+\delta)}$ for $\|\btheta_{\tau}^*-\btheta^*\|$ and of order $\tau^{-(2+2\delta)}$ for $\|\bata^*-\bata_{\tau}^*\|$. As $\tau$ tends to infinity, because the Huber loss becomes the ordinary quadratic loss, the Huber error vanishes as expected. The next theorem studies the smooth error and Huber loss together.

\begin{thm}\label{thm2}
	If Assumptions~(A1)--(A5) hold, then for some $\delta\geq0$ and $h=o(1)$, the minimizer $\bzeta_{\tau,h}^*=((\btheta_{\tau,h}^*)\trans,(\bata_{\tau,h}^*)\trans)\trans$ given in \eqref{eq:smooth_huber} satisfies
	\begin{align*}
		\|\btheta_{\tau,h}^*-\btheta^*\|
		\leq16K_1\left\{\frac{2M_{\delta}+C\|\btheta^*\|^{2+\delta}K_1^{2+3\delta/2}}{K_0(1-\delta_F)(1+\delta)\tau^{1+\delta}}+\frac{\kappa_f\kappa_{k}\|\bbeta^*\|}{K_0(1-\delta_F)}h\right\}
	\end{align*}
	and
	\begin{align*}
		\|\bata_{\tau,h}^*-\bata^*\|
		\leq\frac{64^2K_1^2\kappa_f\kappa_{k}}{\delta_{f_0}\delta_{u_2}\min\{\|\bbeta^*\|,\|\bbeta^*\|^2\}}\left\{\frac{2M_{\delta}+C\|\btheta^*\|^{2+\delta}K_1^{2+3\delta/2}}{K_0(1-\delta_F)(1+\delta)\tau^{1+\delta}}+\frac{\kappa_f\kappa_{k}\|\bbeta^*\|}{K_0(1-\delta_F)}h\right\}^2,
	\end{align*}
	where $C>0$ is a constant.
\end{thm}

\autoref{thm2} reveals that $\|\btheta_{\tau,h}^*-\btheta^*\|$ and $\|\bata_{\tau,h}^*-\bata^*\|$ are associated with both the Huber error and the smooth error. The smoothness parameter $h$ can be of order $\tau^{-(1+\delta)}$, and because $K_h(t)$ approximates $\bone(t\geq0)$ as $h$ tends to zero, the upper bounds in \autoref{thm2} are of the same order as those in \autoref{thm1} when $h\rightarrow0$. Thus, the deviations $\|\btheta_{\tau,h}^*-\btheta^*\|$ and $\|\bata_{\tau,h}^*-\bata^*\|$ are sacrifices in pursuit of robustification and smoothness. The next theorem provides the exponential-type deviations for the baseline and grouping difference parameters and grouping parameter.

\begin{thm}\label{thm3}
	If Assumptions~(A1)--(A5) hold, then for some $\delta$ and any $t>0$, when $h^2n/(p+q+r)\rightarrow\infty$ and $h=o(1)$, the estimator $\hbzeta_{\tau,h}=(\hbtheta_{\tau,h}\trans,\hbata_{\tau,h}\trans)\trans$ given in \eqref{eq:est_huber_em} satisfies, with probability at least $1-21\exp(-t)$,
	\begin{align}\label{eq:th3-theta}
		\begin{split}
			\|\hbtheta_{\tau,h}-\btheta^*\|
			\leq&\frac{32}{K_0(1-\kappa_F)}a(n,\tau)
		\end{split}
	\end{align}
	and
	\begin{align}\label{eq:th3-eta}
		\begin{split}
			\|\hbata_{\tau,h}-\bata^*\|
			\leq&\frac{32}{K_0\|\bbeta^*\|\sqrt{\delta_{f_0}\delta_{k}K_0(1-\kappa_F)}}h^{1/2}a(n,\tau),
		\end{split}
	\end{align}
	where $\nu_0$ is a constant depending on only the constants $\{\kappa_F, \kappa_{f}, \kappa_k,K_1,\|\bxi^*\|\}$, and
	\begin{align*}
		a(n,\tau)=\sqrt{3\nu_0(p+q+r+2t)/n}+\frac{2^{2+\delta}M_{\delta}\{(3+\delta)K_1\|\bbeta^*\|\}^{2+\delta}}{\tau^{1+\delta}}.
	\end{align*}
\end{thm}

With appropriate choice of $\tau$ with $\delta=0$, such as $\tau = O((n/t)^{1/2})$ with $t=\log(n)$, the nonasymptotic property of the sub-Gaussian estimator $\hbtheta_{\tau,h}$ in \autoref{thm3} demonstrates that the deviation $\|\hbtheta_{\tau,h}-\btheta^*\|$ adapts to the sample size, dimension, robustification parameter $\tau$, and moments in pursuit of the optimal tradeoff between bias and robustness,
and the deviation $\|\hbata_{\tau,h}-\bata^*\|$ adapts to the extra smooth parameter $h$ to achieve smoothness. Adaptation to the robustification parameter $\tau$ is caused by pursuing the robustness for linear regression with heavy-tailed errors. The deviations $\|\hbtheta_{\tau,h}-\btheta^*\|$ and $\|\hbata_{\tau,h}-\bata^*\|$ coincide with the smoothed OLS estimator when $\tau\rightarrow\infty$, because $a(n,\tau)=\sqrt{3\nu_0(p+q+r+2t)/n}$ as $\tau\rightarrow\infty$. The next theorem derives the nonasymptotic Bahadur representation of the robust estimators given in \eqref{eq:est_huber_em}.

\begin{thm}
	\label{thm4}
	Let $\bW(\bata) = (\bX\trans,\bZ\trans\bone(\baromega(\bata)\geq0))\trans$ and $\tbW_h(\bata) = (\bX\trans,\bZ\trans K_h(\baromega(\bata))\trans$, where $\baromega(\bata)=U_1+\bU_2\trans\beta$. If the assumptions in \autoref{thm3} are satisfied, then with probability at least $1-32\exp(-t)$ we have
	\begin{align}\label{th4-theta}
		\begin{split}
			\bigg\|\Sigma_W^{1/2}(\hbtheta_{\tau,h}-\btheta^*)-&\Sigma_W^{-1/2}\Pn\left\{\tbW_h(\bata^*)\psi_{\tau}(y-\tbW_h(\bata^*)\trans\btheta^*)\right\}\bigg\|\\
			\leq&\left\{\nu_1\sqrt{(p+q+1+2t)/n}+\nu_2h^{1/2}\right\}a(n,\tau)		
		\end{split}
	\end{align}
	and
	\begin{align}\label{th4-eta}
		\begin{split}
			\bigg\|\Sigma_{U_2}^{1/2}(\hbata_{\tau,h}-\bata^*)-&h\Sigma_{U_2}^{-1/2}\Pn\left\{\bU_2\bZ\trans\bbeta^*K_h(U_1+\bU_2\trans\bata^*)\psi_{\tau}(y-\tbW_h(\bata^*)\trans\btheta^*)\right\}\bigg\|\\
			\leq&\left\{\nu_3\sqrt{(r+2t)/(nh)}+\nu_4\left(h +\tau^{-(1+\delta)}\right)\right\}h^{1/2}a(n,\tau),	
		\end{split}
	\end{align}
	where $a(n,\tau)$ is as defined in \autoref{thm3}, $\nu_1>0$, $\nu_2>0$, $\nu_3>0$, and $\nu_4>0$ are constants depending on only the constants $\{\kappa_F, \kappa_{f}, \kappa_k, K_0, K_1,\|\bxi^*\|\}$, and
	\begin{align*}
		\Sigma_W=\rP\left\{\bW(\bata^*)\bW(\bata^*)\trans\right\},\quad
		\Sigma_{U_2}=K'(0)\rP\left\{f_{\baromega|\bbv}(0)(\bZ\trans\bbeta^*)^2\bU_2\bU_2\trans\right\}.
	\end{align*}
\end{thm}

\autoref{thm4} shows that the remainder of the Bahadur representation of $\|\hbtheta_{\tau,h}-\btheta^*\|$ achieves the rate $h^{1/2}a(n,\tau)$, which is the same as that for $\|\hbata_{\tau,h}-\bata^*\|$. Because of the rate restriction $h^2n/(p+q+r)\rightarrow\infty$ and $h=o(1)$, the remainder of the Bahadur representation in inequality \eqref{th4-eta} does not exhibit subexponential behavior as considered by \cite{Sun2020, Chen2020}. 
This reason is that there is a change plane involved a smooth function in model \eqref{model}. 
To the best of the authors' knowledge, this is the first time that this type of nonasymptotic Bahadur representation
has been reported in the literature, especially for the robust estimator $\hbata_{\tau,h}$ of the grouping parameter, with previous studies reporting only polynomial-type deviation bounds; see \cite{Liu2022a,2021Single}. It is convenient to derive the classical asymptotic results from the Bahadur representation, and Theorems \ref{thm3} and \ref{thm4} show that the robustification parameter $\tau$ and the smoothness parameter $h$ play the same role as bandwidth in constructing classical nonparametric estimators.

\subsection{Implementation}\label{sec:implement}


The theoretical properties in \autoref{properties_est} guarantee that the robust estimation performs well with appropriate choices of the robustification parameter $\tau$ and the smoothness parameter $h$. Because the robustification parameter $\tau$ is treated as a tuning parameter to balance bias and robustness, it is natural to consider using the cross-validation (CV) method to select an appropriate $\tau$ in practice. However, as noted by \citet{Chen2020} and \citet{Wang2021}, because $M_\delta=\E(|\eps|^{2+\delta})$ is typically unknown in practice, its empirical OLS estimator $\hat{M}_\delta=(n-p-q)^{-1}\sum_{i=1}^{n}(y_i-\bX_i\trans\hat{\balpha}_{\tau,h}-\bZ_i\trans\hat{\bbeta}_{\tau,h}\bone(\bU_i\trans\hat{\gamma}_{\tau,h}\geq0))^{2+\delta}$ is poor when the errors are heavy-tailed. Instead, there are two good alternatives \citep{Chen2020}: (i) an adaptive technique based on Lepski's method \cite{Lepskii1992} and (ii) a Huber-type method by solving a so-called censored equation \cite{Hahn1990}; see \cite{Chen2020,Wang2021} for details. For the smoothness parameter $h$, the CV method is always a natural choice for selecting an appropriate one. Alternatively, from the theoretical conditions on the smoothness parameter $h$, a rule of thumb $h_n=c_h\hat{\sigma}_u\log(n)/\sqrt{n}$ suggested by \citet{2007A}, \citet{2021Multithreshold}, and \citet{2021Single} is recommended by considering the computation reduction, where $c_h$ is a constant and $\hat{\sigma}_u=\sqrt{(n-r)^{-1}\sum_{i=1}^{n}(U_{1i}+\bU_{2i}\trans\hbata^{ols})^2}$ is the estimated standard deviation of $\bU\trans\bgamma$, with $(\hbalpha^{ols},\hbbeta^{ols},\hbata^{ols})$ being the estimator of $(\balpha,\bbeta,\bata)$ by using ordinary quadratic loss.

Attention now turns to the implementation for subgroup-classifier learning, with the parameters $\btheta = (\balpha\trans, \bbeta\trans)\trans$ and $\bata$ estimated iteratively as follows. Let $\ell_\tau(\balpha,\bbeta, \bata)$ be the loss function for model \eqref{model}, i.e.,
\begin{align*}
	\ell_\tau(\balpha,\bbeta, \bata) = \sum_{i=1}^nL_{\tau}\left(y_i-\bX_i\trans\balpha - \bZ_{i}\trans\bbeta\bone(U_{1i}+\bU_{2i}\trans\bata\geq 0)\right),
\end{align*}
and let the smoothed loss function be
\begin{align}\label{approx_loss_multi}
	\tilde{\ell}_{\tau,h}(\balpha,\bbeta, \bata) = \sum_{i=1}^nL_{\tau}\left(y_i-\bX_i\trans\balpha - \bZ_{i}\trans\bbeta K_h(U_{1i}+\bU_{2i}\trans\bata\geq 0)\right).
\end{align}
For given $\bata^{(k)}$, one obtains $\balpha^{(k+1)}$ and $\bbeta^{(k+1)}$ by minimizing the smoothed loss function
\begin{align*}
	(\balpha^{(k+1)}, \bbeta^{(k+1)}) = \argmin_{\balpha\in\Theta_{\alpha},\bbeta\in\Theta_{\beta}}\tilde{\ell}_{\tau,h}(\balpha,\bbeta, \bata^{(k)}),
\end{align*}
and for given $\balpha^{(k+1)}$ and $\bbeta^{(k+1)}$, one estimates $\bata^{(k+1)}$ by
\begin{align*}
	\bata^{(k+1)} = \argmin_{\bata\in\Theta_{\eta}}\tilde{\ell}_{\tau,h}(\balpha^{(k+1)},\bbeta^{(k+1)}, \bata).
\end{align*}
Iterating these two minimizers leads to the desired robust estimators. These are summarized with the multiplier bootstrap calibration in Algorithm 1 in Appendix A of the Supplementary Material, which provides the strategy for estimating the confidence intervals for the estimators $\hbalpha$, $\hbbeta$, and $\hbata$.

Note that herein, $\balpha^{(k+1)}$ and $\bbeta^{(k+1)}$ for given $\bata^{(k)}$ are obtained by a robust data-adaptive method proposed by \citet{Wang2021}. Specifically, $\btheta = (\balpha\trans,\bbeta\trans)\trans$ is estimated and $\tau$ is calibrated simultaneously by solving the following system of equations:
\begin{align*}
	\left\{
	\begin{array}{l}
		\sum_{i=1}^{n}\psi_{\tau}(y_i-\bW_i\trans\btheta)\bW_i=0,\\
		(\tau^2 n)^{-1}\sum_{i=1}^{n}\min\{(y_i-\bW_i\trans\btheta)^2,\tau^2\}-n^{-1}(d+z)=0,
	\end{array}
	\right.
\end{align*}
where $d=p+q-1$, $\bW_i=(\bX_i\trans,\bZ_i\trans\bone(U_{1i}+\bU_{2i}\trans\bata^{(k)}\geq0))\trans$, and $z=\log(n)$ as suggested by \citet{Wang2021}. The initial values $\btheta^{(0)}=\btheta^{(ols)}$ and $\tau^{(0)}=\hat{\sigma}_\eps\sqrt{n/(d+z)}$ are set using the ordinary quadratic loss, where $\hat{\sigma}_\eps^2=(n-p-q-r)^{-1}\sum_{i=1}^{n}(y-\bX_i\trans\hbalpha^{ols}-\bZ_i\trans\hbbeta^{ols}\bone(U_{1i}+\bU_{2i}\trans\hbata^{ols}\geq0))^2$, with $(\hbalpha^{ols},\hbbeta^{ols},\bata^{ols})$ being the estimator of $(\balpha,\bbeta,\bata)$.


\section{Robust subgroup testing}\label{sec:test}

Before learning the subgroup classifier, it is of interest to test for the existence of subgroups, which guarantees avoidance of the identifiability problem of $\bata$. This section considers the test problem \eqref{test}. Recall the loss function in \eqref{eq:est_huber}, i.e.,
\begin{align}\label{eq:loss_indicator}
	L_{\tau}(y-\bX\trans\balpha-\bZ\trans\bbeta\bone(\bU\trans\bgamma\geq0)),
\end{align}
the derivative of which with respect to $\bbeta$ under the alternative hypothesis is
\begin{align*}
	\varphi(\bV, \balpha, \bbeta, \bgamma)=\bZ\bone(\bU\trans\bgamma\geq0)\psi_{\tau}(y-\bX\trans\balpha-\bZ\trans\bbeta\bone(\bU\trans\bgamma\geq0) ),
\end{align*}
and with respect to $\balpha$ under the null hypothesis is
\begin{align*}
	\varphi_0(\bV, \balpha)=\bX\psi_{\tau}(y-\bX\trans\balpha),
\end{align*}
where $\psi_{\tau}(u)$ is the first derivative of the Huber loss \eqref{eq:loss_huber}, defined as $\psi_{\tau}(u)=\sgn(u)\min\{|u|,\tau\}$.

\subsection{Robust estimation under null hypothesis}

Under the null hypothesis, model \eqref{model} reduces to the ordinary linear regression model with heavy-tailed errors, i.e.,
\begin{align}\label{model_H0}
	y_i=\bX_i\trans\balpha + \eps_i.
\end{align}
Parametric estimation in model \eqref{model_H0} is well-studied in the literature; see \cite{Huber1964, Huber1973, Fan2017, Sun2020, Wang2021, Han2021, Chen2020, Zhou2018}, among others. Let $\hbalpha_{\tau}$ be the estimate of $\balpha_{\tau}$ under the null hypothesis, i.e.,
\begin{align}\label{eq:opt_H0}
	\hbalpha_{\tau} = \argmin_{\balpha\in\Theta_{\alpha}}\Pn L_{\tau}(y-\bX\trans\balpha).
\end{align}

Before establishing asymptotic properties for the robust estimator under the null hypothesis and the test statistic, the following required assumptions are made.

\begin{enumerate}[({B}1)]
	\item The random vector $\bX$ is sub-Gaussian, $J=\{\rP[\bX\bX\trans]\}^{-1}\in\bR^{p\times p}$ is a finite and positive definite deterministic matrix, and there is a universal constant $K_1>0$ satisfying $\|\bX\|_{\psi_2}\leq K_1$, $\|\bZ\|_{\psi_2}\leq K_1$.
	\item The error variable $\eps$ is independent of $(\bX\trans,\bZ\trans,\bU\trans)\trans$ and satisfies $\rP(\eps)=0$ and $\rP(|\eps|^{2+\delta})=M_{\delta}<\infty$ with $\delta\geq0$. 
	\item $0<\rP[\bone(\bU\trans\bgamma\geq0)]<1$ for any $\bgamma\in\Theta_{\gamma}\subseteq\mathbb{R}^r$.
\end{enumerate}

\begin{remark}
	Assumption~(B1) is the moment condition of covariates for establishing the nonasymptotic properties under the null hypothesis and deriving the asymptotic distributions. Assumption~(B2) is the same as Assumption~(A2), and Assumption~(B3) is the same as Assumption~(A3).
\end{remark}

\begin{lem}\label{thm0}
	(\citet{Chen2020}) If Assumptions~(B1)--(B3) hold, then for any $t>0$ and $v\geq v_{2+\delta}^{1/(2+\delta)}$, the estimator $\hbalpha_{\tau}$ given in \eqref{eq:opt_H0} with $\tau=v(\frac{n}{p+t})^{1/(2+\delta)}$ satisfies
	\begin{align}
		\begin{split}
			\rP\left\{\left\|J^{-1/2}(\hbalpha_{\tau}-\balpha^*)\right\|\geq c_1v\sqrt{\frac{p+t}{n}}\right\}\leq& 2e^t\\
			\mbox{and}~~\rP\left\{\left\|J^{-1/2}(\hbalpha_{\tau}-\balpha^*)-J^{-1/2}\Pn\bX\psi_{\tau}(\eps)\right\|\geq c_2v\sqrt{\frac{p+t}{n}}\right\}\leq& 2e^t
		\end{split}
	\end{align}
	as along as $n\geq c_3(p+t)$, where $c_1$--$c_3$ are constants depending on only $K_1$.
\end{lem}

\subsection{Robust test statistic}\label{sec:rwast}

As discussed by \citet{Liu2022a}, for any known $\bgamma\in\Theta_{\gamma}$, it is natural to consider an SST statistic for testing $\bbeta=\bzero$, i.e.,
\begin{align}\label{score_test0}
	\tT_n(\bgamma) = n^{-1}\|\Pn \varphi(\bV, \hbalpha_{\tau}, \bzero, \bgamma)\|^2_{\tV(\bgamma)^{-1}},
\end{align}
where $\hbalpha_{\tau}$ is given in \eqref{eq:opt_H0} and $\tV(\bgamma)=\Pn\{\varphi(\bV, \hbalpha_{\tau}, \bzero, \bgamma) - \hat{G}(\bgamma)\hat{J}\varphi_0(\bV, \hbalpha_{\tau})\}^{\otimes2}$. Here, $\hG(\bgamma)$ and $\hJ$ are consistent estimators of $G(\bgamma)$ and $J$, respectively, where 
$J$ is as defined in Assumption~(B1) and
\begin{align*}
	G(\bgamma)=\rP\{\bZ\bX\trans\bone(\bU\trans\bgamma\geq0)\bone(|y-\bX\trans\balpha^*|\leq\tau)\}\in\mathbb{R}^{q\times p}.
\end{align*}

\begin{lem}\label{lem1}
	If Assumptions~{(B1)--(B3)} hold, then for any fixed $\bgamma\in \Theta_\gamma$, $\tT_n(\bgamma)$ converges in distribution to a $\chi^2$ one with $q$ degrees of freedom under $H_0$ as $n \rightarrow \infty$.
\end{lem}

Although there is an unknown parameter $\bgamma$ that prevents $\tT_n(\bgamma)$ from being used directly in practice, Lemma~\ref{lem1} reveals essentially that the asymptotic distribution of $\tT_n(\bgamma)$ is free of the nuisance parameter $\bgamma$. Thus, the supremum and the weighted average of $\tT_n(\bgamma)$ over $\bgamma$ should guarantee the correct type-I errors, which motivates constructing the robust test statistic based on the weighted average of $\tT_n(\bgamma)$ over the parametric space $\Theta_{\gamma}$.

\citet{2017Change} studied the supremum of the SST statistic $\tT_n(\bgamma)$ (SST) over the grouping parameter $\bgamma$ for a semiparametric model, i.e.,
\begin{align}\label{test_sup}
	\tT_n = \sup_{\bgamma\in\Theta_{\gamma}} \left\{n^{-1}\|\Pn\varphi(\bV, \hbalpha_{\tau}, \bzero, \bgamma)\|^2_{\tV(\bgamma)^{-1}}\right\}.
\end{align}
The test statistic $\tT_n$ has been investigated widely in the literature; see \cite{1994Optimal,1995Admissibility, 1977Hypothesis, 2009On, Shen2020, Liu2022a}. It is easy to extent the SST to model \eqref{model} with heavy-tailed errors according to the Bahadur representation in \autoref{thm4}.


When the dimension of the parametric space $\Theta_{\gamma}$ is large, searching for the supremum value over $\Theta_{\gamma}$ may cause $\tT_n$ to lose power in practice and is time-consuming computationally. To avoid these drawbacks, proposed in this section is a robust test procedure that is a type of WAST statistic first introduced by \citet{Liu2022a}.

Proposed herein is RWAST, i.e.,
\begin{align}\label{statistic}
	T_n = \frac{1}{n(n-1)}\sum_{i\neq j}\omega_{ij}\bZ_i\trans\bZ_j\psi_{\tau}(y_i-\bX_i\trans\hbalpha_{\tau})\psi_{\tau}(y_j-\bX_j\trans\hbalpha_{\tau}),
\end{align}
where
\begin{align}\label{wij}
\omega_{ij}= \frac{1}{4}+\frac{1}{2\pi}\arctan\left(\frac{\varrho_{ij}}{\sqrt{1-\varrho_{ij}^2}}\right) \quad \mbox{if}~ i\neq j,
\end{align}
and $\varrho_{ij} = \bU_i\trans\bU_j(\|\bU_i\|\|\bU_j\|)^{-1}$. As noted by \citet{Liu2022a}, there is a Bayesian explanation for the weight $\omega_{ij}$. In fact, Lemma~D1 in the Supplementary Material shows that
\begin{align*}
	\omega_{ij} = \int_{\bgamma\in\Theta_{\gamma}}\bone(\bU_i\trans\bgamma\geq0) \bone(\bU_j\trans\bgamma\geq0)w(\bgamma)d\bgamma,
\end{align*}
where $w(\bgamma)$ is the standard multi-Gaussian density and can be chosen as another weight satisfying $w(\bgamma)\geq0$ for all $\bgamma\in\Theta_{\gamma}$ and $\int_{\bgamma\in\Theta_{\gamma}} w(\bgamma)d\bgamma=1$. 
In the Bayesian motivation, the grouping parameter $\bgamma$ has a prior with density $w(\bgamma)$. Because the goal herein is to test for the existence of subgroups instead of estimating the grouping parameter, the difference is that there is no requirement for the posterior distribution.


The choice of the weight affect the computation of the test statistic because of the numerical integration over $\bR^q$. 
Thus, taking the weight as the standard multi-Gaussian density offers good performance in practice, as illustrated in the simulation studies in \autoref{sec:sim} and the case studies in \autoref{sec:case}. To illustrate the performance in robust regression, numerical studies were conducted to investigate the sensitivity of the weight's choice by comparing $\omega_{ij}$ with the closed form in \eqref{wij} and the approximated $\omega_{ij}$ in (E.3) of Appendix~E.3 in the Supplementary Material. From the numerical results, compared with the approximated $\omega_{ij}$, the test statistic with $\omega_{ij}$ in \eqref{wij} has higher power uniformly and takes only 10\% of the time computationally when $N=10\,000$. This is a strong recommendation to use the closed-form RWAST in \eqref{wij}.


To establish the asymptotic distribution of RWAST, additional notation is introduced below. Denote the kernel of a U-statistic under the null hypothesis by
\begin{align}\label{kernel0}
\begin{split}
	h(\bV_i, \bV_j)
	=& \omega_{ij}\bZ_i\trans\bZ_j\psi_{\tau}(y_i-\bX_i\trans\balpha^* )\psi_{\tau}(y_j-\bX_j\trans\balpha^*)\\
	& + \varphi_0(\bV_i, \balpha^*)\trans K_{j}+ K_{i}\trans \varphi_0(\bV_j, \balpha^*) + \varphi_0(\bV_i, \balpha^*)\trans H\varphi_0(\bV_j, \balpha^*),
\end{split}
\end{align}
where $\varphi_0(\bV,\balpha)=\bX\psi_{\tau}(y-\bX\trans\balpha)$, and
\begin{align*}
H=\int_{\bgamma\in\Theta_{\gamma}}J\trans G(\bgamma)\trans G(\bgamma)J w(\bgamma)d\bgamma ~\mbox{and}~
K_i = \int_{\bgamma\in\Theta_{\gamma}}J\trans G(\bgamma)\trans\varphi(\bV_i, \balpha^*, \bzero, \bgamma) w(\bgamma)d\bgamma
\end{align*}
with $\varphi(\bV,\balpha,\bbeta,\bgamma)=\bZ\bone(\bU\trans\bgamma\geq0)\psi_{\tau}(y-\bX\trans\balpha-\bZ\trans\bbeta\bone(\bU\trans\bgamma\geq0))$.

\begin{thm}\label{thm11}
If Assumptions~{(B1)--(B3)} hold, then under the null hypothesis, we have
\begin{align*}
	nT_n -\mu_0 \lkonv \nu,
\end{align*}
where $\lkonv$ denotes convergence in distribution, $\mu_0=n\{\rP\psi_{\tau}(\eps)\}^2\int_{\bgamma\in\Theta_\gamma}\{\rP\bZ\bone(\bU\trans\bgamma\geq0)\}^2w(\bgamma)d\bgamma+\rP[\varphi_0(\bV, \balpha^*)\trans H\varphi_0(\bV, \balpha^*)]+2\rP[\varphi_0(\bV_1, \balpha^*)\trans K_{1}]$, $\nu$ is a random variable of the form $\nu=\sum_{j=1}^{\infty}\lambda_{j}(\chi^2_{1j}-1)$, and $\chi^2_{11}, \chi^2_{12}, \cdots$ are independent $\chi^2_{1}$ variables, i.e., $\nu$ has the characteristic function
\begin{align*}
	\rP\left[e^{it\nu}\right]=\prod_{j=1}^{\infty}(1-2it\lambda_{j})^{-\frac{1}{2}}e^{-it\lambda_{j}}.
\end{align*}
Here, $i=\sqrt{-1}$ is the imaginary unit, and $\{\lambda_{j}\}$ are the eigenvalues of the kernel $h(\bv_1, \bv_2)$ under $f(\bv, \balpha^*, \bzero, \bgamma^*)$, i.e., they are the solutions of $\lambda_{j}g_{j}(\bv_2)=\int_{0}^{\infty}h(\bv_1, \bv_2)g_{j}(\bv_1)\\ f(\bv_1, \balpha^*, \bzero, \bgamma^*)d\bv_1$ for nonzero $g_{j}$, where $f(\bv, \balpha, \bbeta, \bgamma)$ is the density of $\bV$. 
\end{thm}


Investigated next is the power performance of the proposed test statistic under two types of alternative hypotheses under which subgroups exist. Considered first is the global alternative denoted by $H_{1g}$: $\bbeta=\bxi$, where $\bxi\in\Theta_{\bbeta}\backslash\{\bzero\}$ is fixed. \autoref{thm120} provides the asymptotic distribution of the test statistic $T_n$ under the global alternative.

\begin{thm}\label{thm120}
If Assumptions~{(B1)--(B3)} hold, then under the global alternative $H_{1g}$, we have
\begin{align*}
	\sqrt{n}(T_n - \mu_1) \lkonv \mathcal{N}(0, \sigma^2_{\xi}),
\end{align*}
where $\mu_1 = \rP[h(\bV_1, \bV_2)]$ and $\sigma^2_{\xi}=4\Var(\rP[h(\bV_1, \bV_2)|\bV_1])$.
\end{thm}

To derive the asymptotic distribution under the local alternative hypothesis, denoted by $H_{1l}: \bbeta = n^{-1/2}\bxi$, additional assumptions are required, where $\bxi\in\Theta_{\beta}$ is a fixed vector.

\begin{enumerate}[({B}4)]
\item 
There is a positive function $b(\bv, \bxi)$ of $\bv$ relying on $\balpha^*$, $\bgamma^*$ such that
\begin{align*}
	\left|\bxi\trans\frac{\partial f(\bv; \balpha^*, r_n\bxi, \bgamma^*)\partial\bbeta}{f(\bv; \balpha^*, \bzero, \bgamma^*)}\right|\leq b(\bv, \bxi),
\end{align*}
and $\rP[b(\bV, \bxi)^2]$ and $\rP[\phi_k(\bV)^2b(\bV, \bxi)]$ for all $k=1,\cdots,$ are bounded by $C_{\bxi}$, where $\bxi\in\Theta_{\beta}$, $r_n=o(1)$, $C_{\bxi}>0$ is a constant relying on $\bxi$, $\phi_k(\cdot)$ is as defined in \autoref{thm11}, and $\bV$ is generated from the null distribution with density $f(\bv; \balpha^*, \bzero, \bgamma^*)$.
\end{enumerate}

\begin{thm}\label{thm12}
If Assumptions~{(B1)--(B4)} hold, then under the local alternative hypothesis $H_{1l}$, i.e., $\bbeta = n^{-1/2}\bxi$ with a fixed vector $\bxi\in\Theta_{\beta}$, we have
\begin{align*}
	nT_n-\mu_0 \lkonv \nu,
\end{align*}
where $\mu_0$ is as defined in \autoref{thm11}, $\nu$ is a random variable of the form $\nu=\sum_{j=1}^{\infty}\lambda_{j}(\chi^2_{1j}(\mu_{aj})-1)$, and $\chi^2_{11}(\mu_{a1}), \chi^2_{12}(\mu_{a2}), \cdots$ are independent noncentral $\chi^2_{1}$ variables, i.e., $\nu$ has the characteristic function
\begin{align*}
	\rP\left[e^{it\nu}\right]=\prod_{j=1}^{\infty}(1-2it\lambda_{j})^{-\frac{1}{2}}
	\exp\left(-it\lambda_j+\frac{it\lambda_{j}\mu_{aj}}{1-2it\lambda_j}\right).
\end{align*}
Here, $\{\lambda_{j}\}$ are the eigenvalues of the kernel $h(\bv_1, \bv_2)$ defined in \eqref{kernel0} under $f(\bv, \balpha^*, \bzero, \bgamma^*)$, i.e., they are the solutions of $\lambda_{j}g_{j}(\bv_2)=\int_{0}^{\infty}h(\bv_1, \bv_2)g_{j}(\bv_1)f(\bv_1, \balpha^*, \bzero, \bgamma^*)d\bv_1$ for nonzero $g_{j}$, and each noncentrality parameter of $\chi^2_{1j}(\mu_{aj})$ is
\begin{align*}
	\mu_{aj} = \rP\left[\phi_j(\bV_{0})\bxi\trans\partial \log(f(\bV_{0}, \balpha^*, \bzero, \bgamma^*))/\partial\bbeta\right], \quad j=1, 2, \cdots,
\end{align*}
where $\{\phi_j(\bv)\}$ denotes orthonormal eigenfunctions corresponding to the eigenvalues $\{\lambda_j\}$, and $\bV_0$ 
is generated from the null distribution $f(\bv, \balpha^*, \bzero, \bgamma^*)$.
\end{thm}


Denote by $F_\nu$ the cumulative distribution function of $\nu$. It follows from \autoref{thm12} that the power function of $nT_n-\mu_0$ can be approximated theoretically by $F_\nu$, and the proof of \autoref{thm12} shows that $0< \rP h(\bV_1, \bV_2)=\sum_{j=1}^{\infty}\lambda_j\mu_{aj}+o(1)$ under the local alternative hypothesis. The additional mean $\mu_1$ under $H_{1g}$ (or $\sum_{j=1}^{\infty}\lambda_j\mu_{aj}$ under $H_{1l}$) can be viewed as a measure of the difference between $H_0$ and $H_{1g}$ (or $H_{1l}$). $F_\nu$ is difficult to use in practice because it is not common. In Appendix B of the Supplementary Material, a novel bootstrap method is recommended for calculating the critical value or $\pvalue$, of which the asymptotic consistency is established in Theorem B1 in the Supplementary Material.

\section{Simulation studies}\label{sec:sim}

Consider the change-plane model \eqref{model}
\begin{align*}
y_i=\bX_i\trans\balpha+\bX_i\trans\bbeta\bone(\bU_i\trans\bgamma\geq0) + \eps_i,
\end{align*}
where $X_{1i}=1$ and $U_{1i}=1$, and $(X_{2i},\cdots,X_{pi})\trans$ and $(U_{2i},\cdots,U_{ri})\trans$ are generated independently from multivariate normal distributions $N(\bzero_{p-1},\sqrt{2}\bI_{p-1})$ and $N(\bzero_{r-1},\sqrt{2}\bI_{r-1})$, respectively. The error $\eps_i$ is generated from following eight distributions:
(i) $N(0,\sqrt{2})$;
(ii) $t_2$;
(iii) Pareto distribution $Par(2, 1)$ with shape parameter 2 and scale parameter 1;
(iv) Weibull distribution $Weib(0.75, 0.75)$ with shape parameter 0.75 and scale parameter 0.75.
For the limit of space, we put other four error's distributions in Appendix E of the Supplementary Material: (v) Gaussian mixture;
(vi) mixture of $t_2$ and Weibull $Weib(0.75, 0.75)$;
(vii) mixture of Pareto $Par(2, 1)$ and $N(0,\sqrt{2})$; and
(viii) mixture of lognormal $\exp(N(0, 1))$ and $N(0,\sqrt{2})$.
$(\gamma_2, \cdots,\gamma_q)\trans = (1,2,\cdots,2)\trans$ is set under $H_1$, and $\gamma_1$ is chosen as the negative of 35\% percentile of $U_2\gamma_2+\cdots+U_q\gamma_r$, which means that $\bU\trans\bgamma$ divides the population into two groups with 65\% and 35\% observations, respectively. To save space, we only present simulation results of robust subgroup-classifier learning, and we put performance of robust subgroup testing in Appendix E2 of the Supplementary Material.


The settings used herein are $\balpha = (5,0.5,\cdots,0.5)\in \mathbb{R}^{p} \quad \mbox{and} \quad \bbeta = (0.5,\cdots,0.5)\in\mathbb{R}^{q},$ with the sigmoid function $K(u) = \{1+\exp(-u)\}^{-1}$ chosen as the smooth function. Finite-sample studies were performed for different smooth functions $K(u)$ (such as $K(u)=\Phi(u)$ and $K(u)=\Phi(u)+u\phi(u)$), but the results were similar and so are omitted here. For comparison, three strategies are considered:
(i) the proposed adaptive procedure (AHu),
(ii) the classic Huber method (Hub), with the robustness parameter $\tau$ selected by $\tau_0\mbox{median}\{\by-\mbox{median}(\by)\}/\Phi^{-1}(0.75)$ with $\tau_0=1.345$ and $\by=(y_1,\cdots,y_n)\trans$, as suggested in \cite{Wang2021}), and
(iii) the estimation method based on the ordinary quadratic loss considered in \cite{2021Multithreshold} (OLS).
Here, the subscript $n$ in AHU$_n$, Hub$_n$ and OLS$_n$ stands for the sample size with $n=200, 400, 600$.

\begin{figure}[!th]
\begin{center}
	\includegraphics[height=2.7in,width=1.5in, angle=-90]{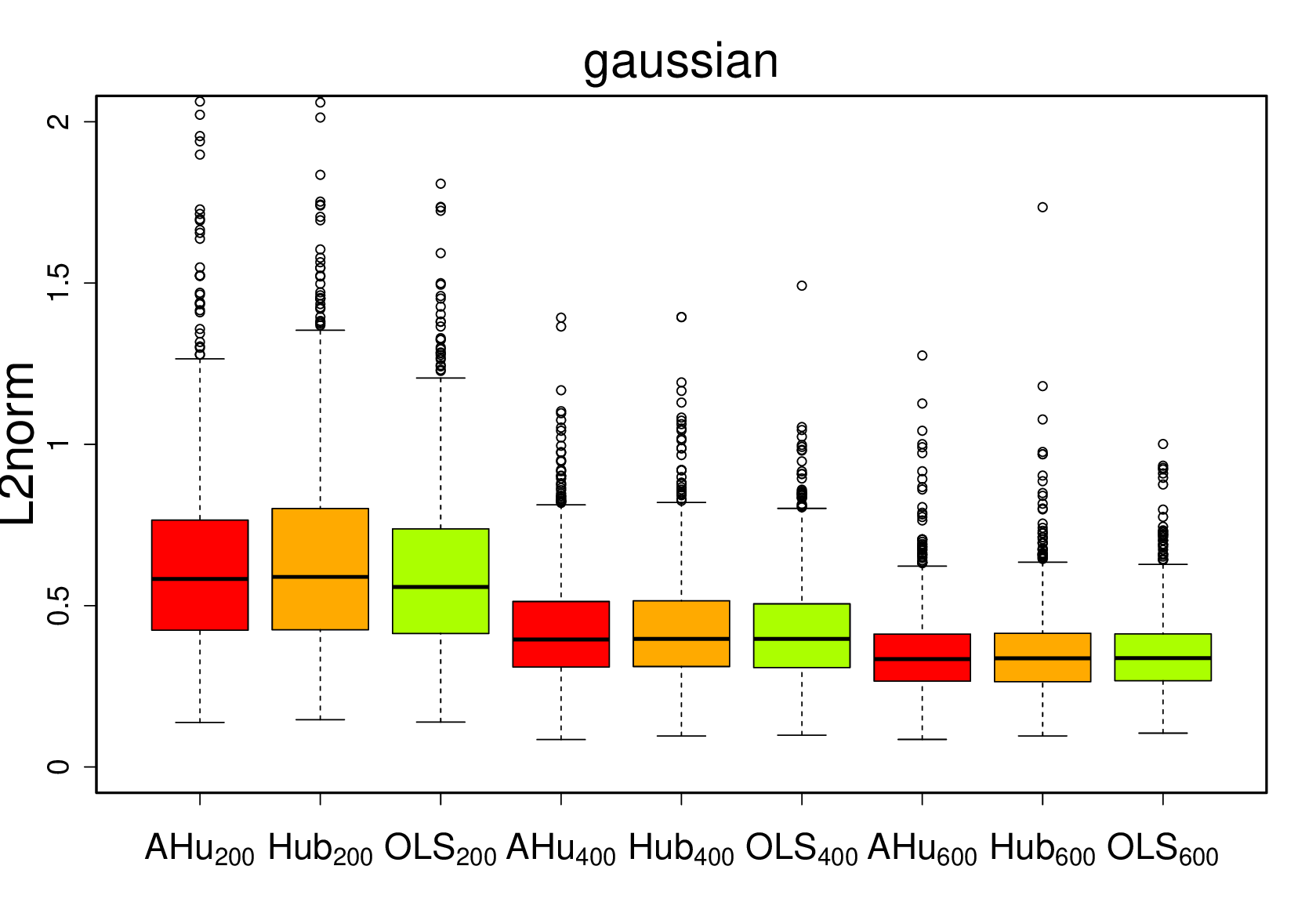}
	\includegraphics[height=2.7in,width=1.5in, angle=-90]{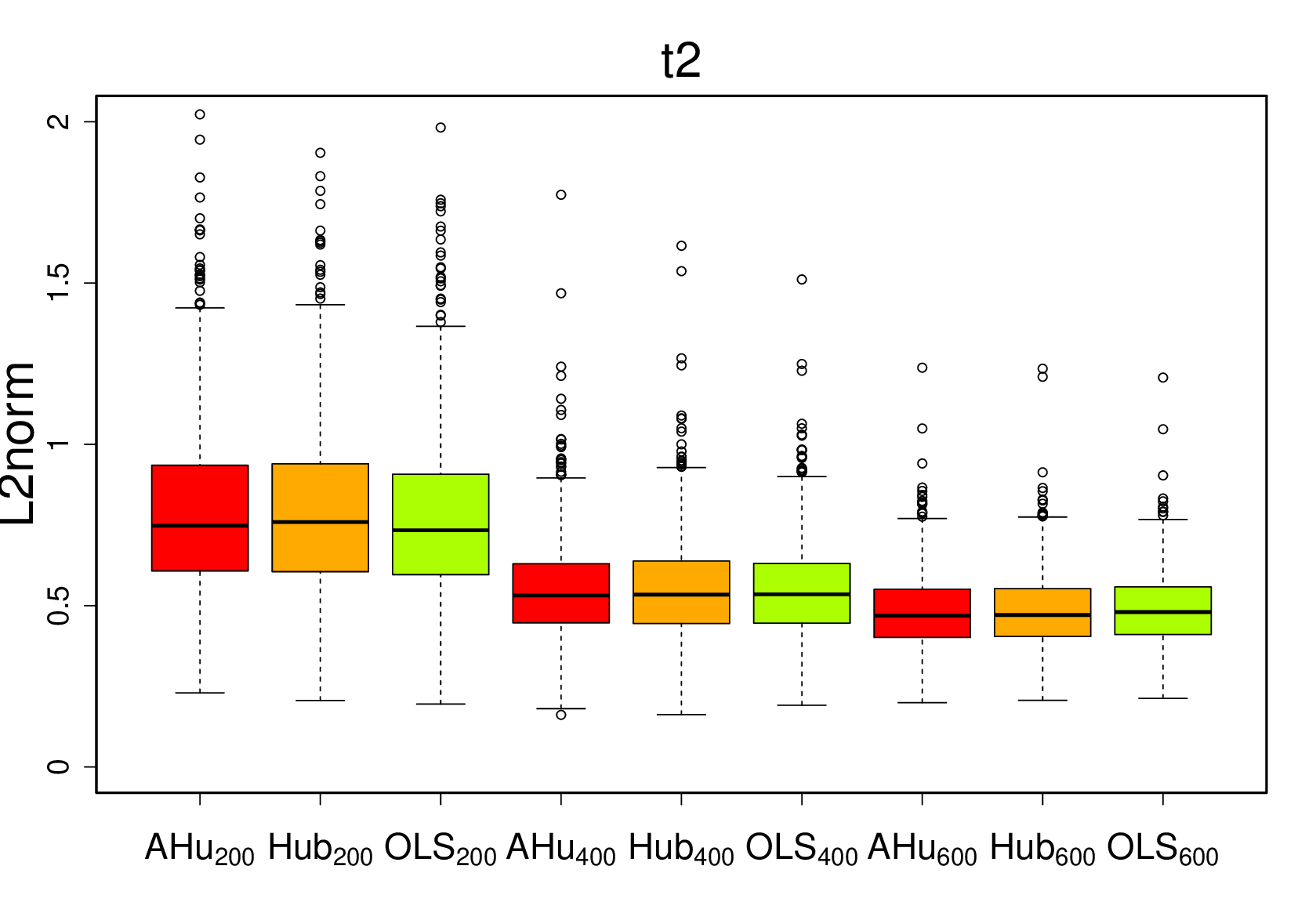}
	\includegraphics[height=2.7in,width=1.5in, angle=-90]{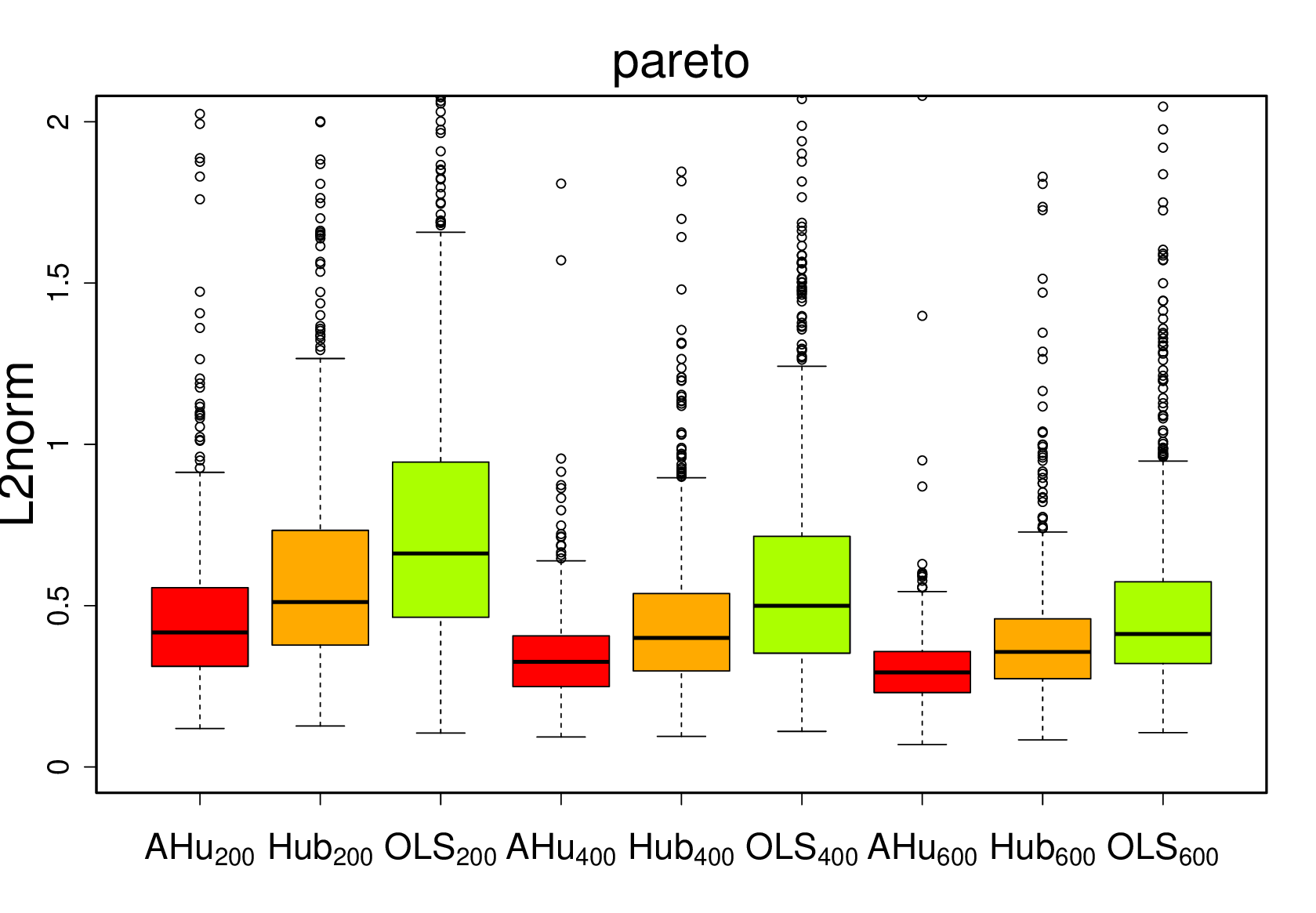}
	\includegraphics[height=2.7in,width=1.5in, angle=-90]{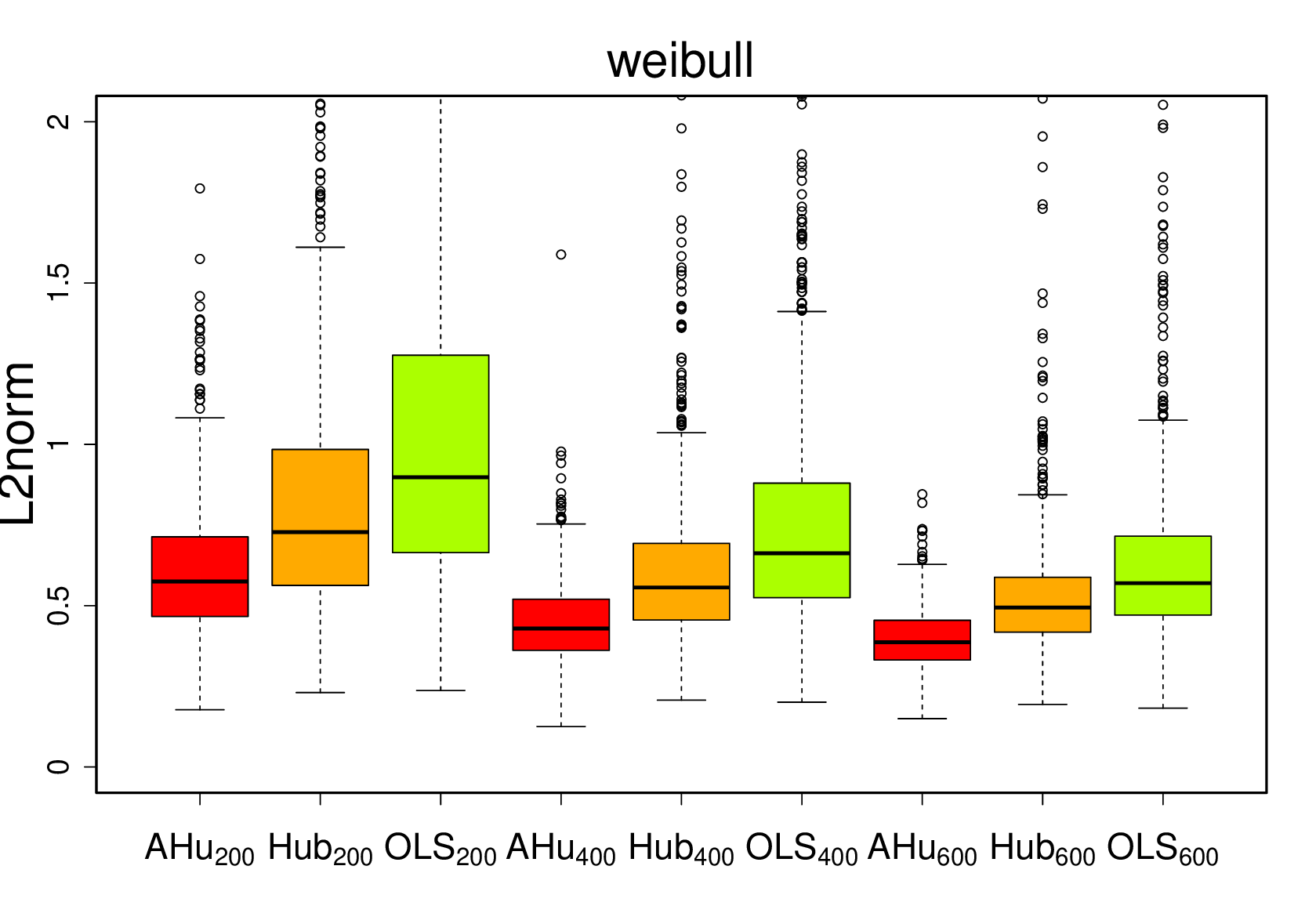}
	\caption{\it $L_2$-norm of estimation errors of parameter $\btheta = (\balpha\trans, \bbeta\trans)\trans$ with four different error distributions. Here, $(p, q, r)=(3, 3, 3)$.}
	\label{fig_L2_qr33}
\end{center}
\end{figure}

\begin{figure}[!th]
\begin{center}
	\includegraphics[height=2.7in,width=1.5in, angle=-90]{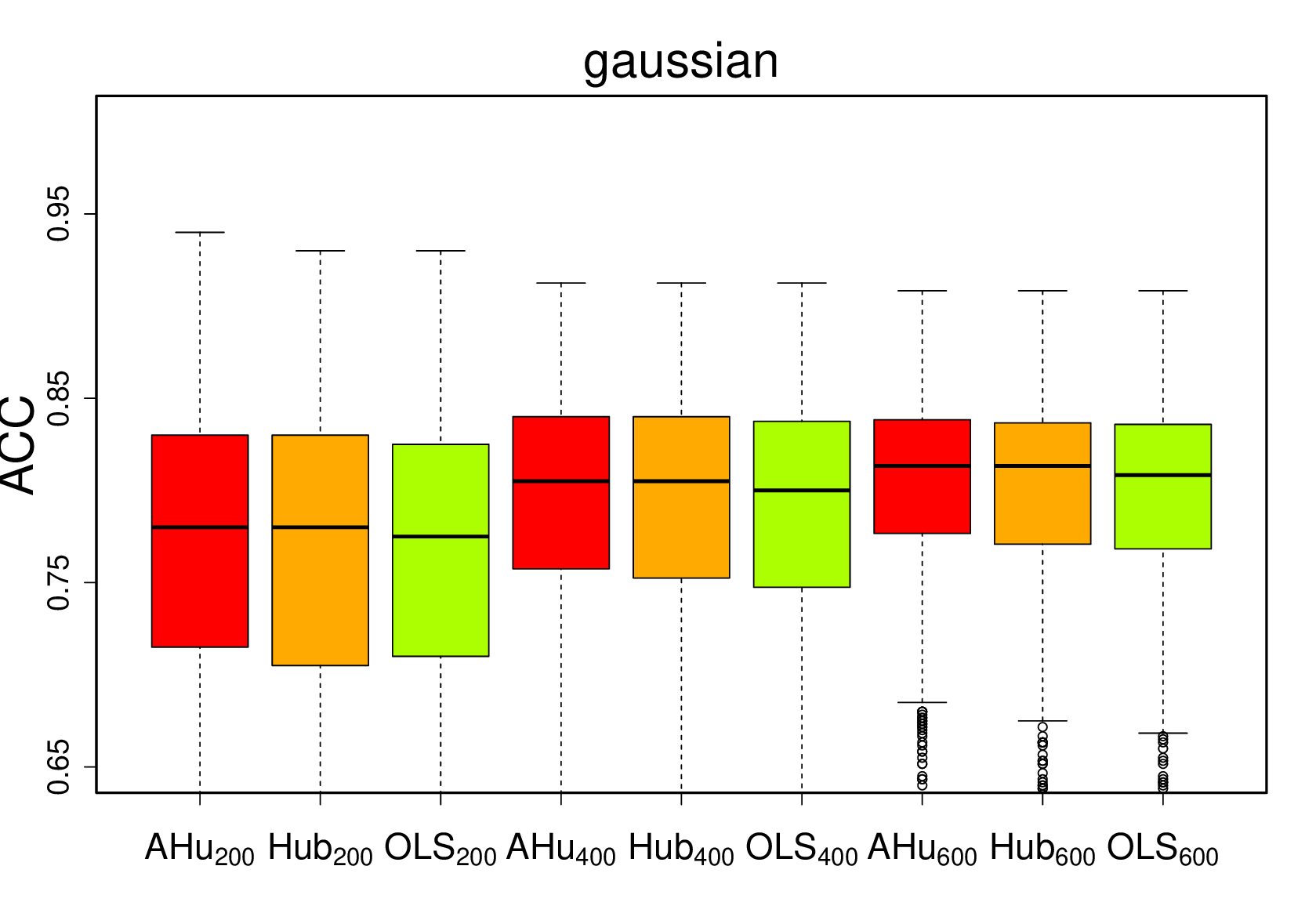}
	\includegraphics[height=2.7in,width=1.5in, angle=-90]{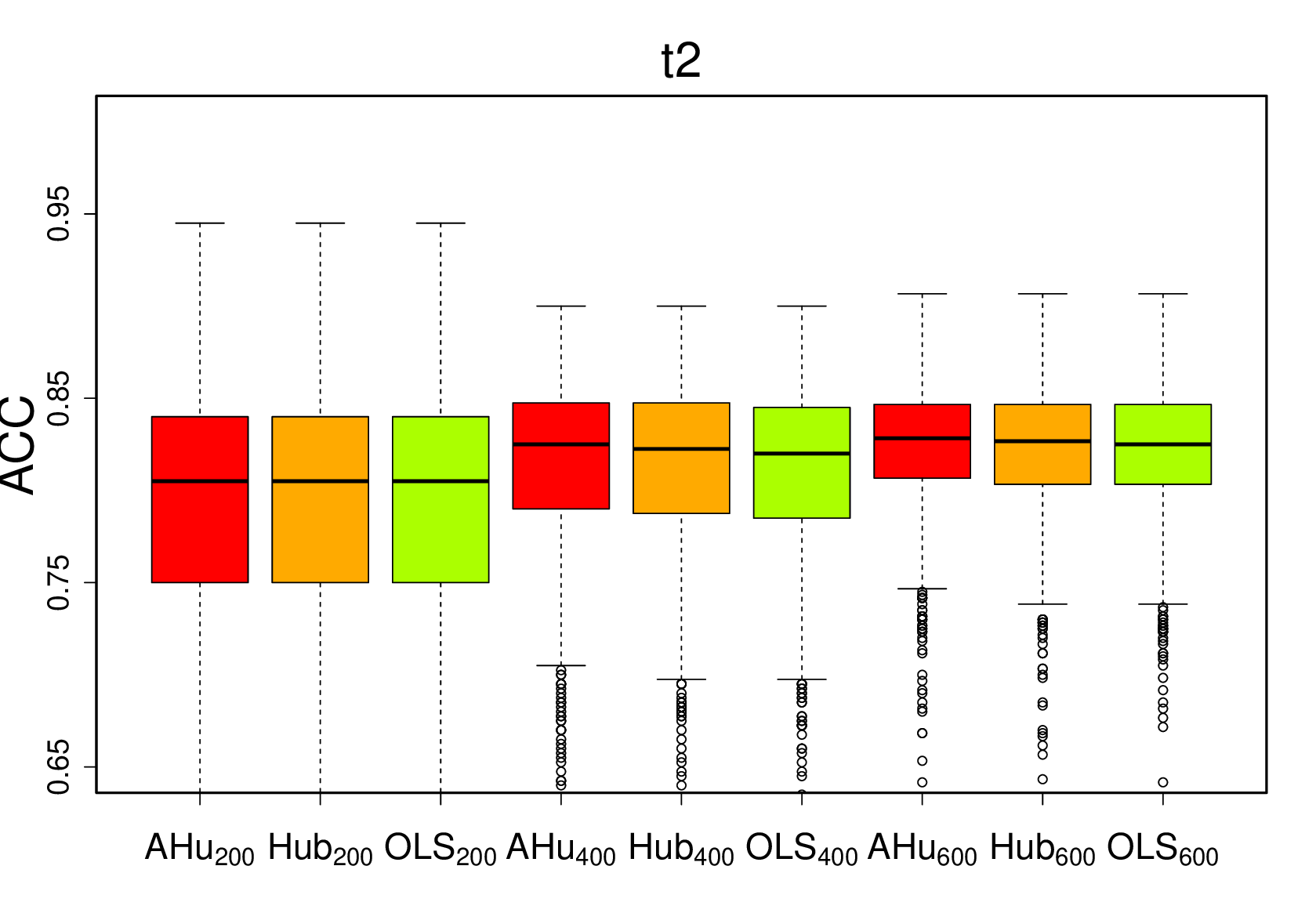}
	\includegraphics[height=2.7in,width=1.5in, angle=-90]{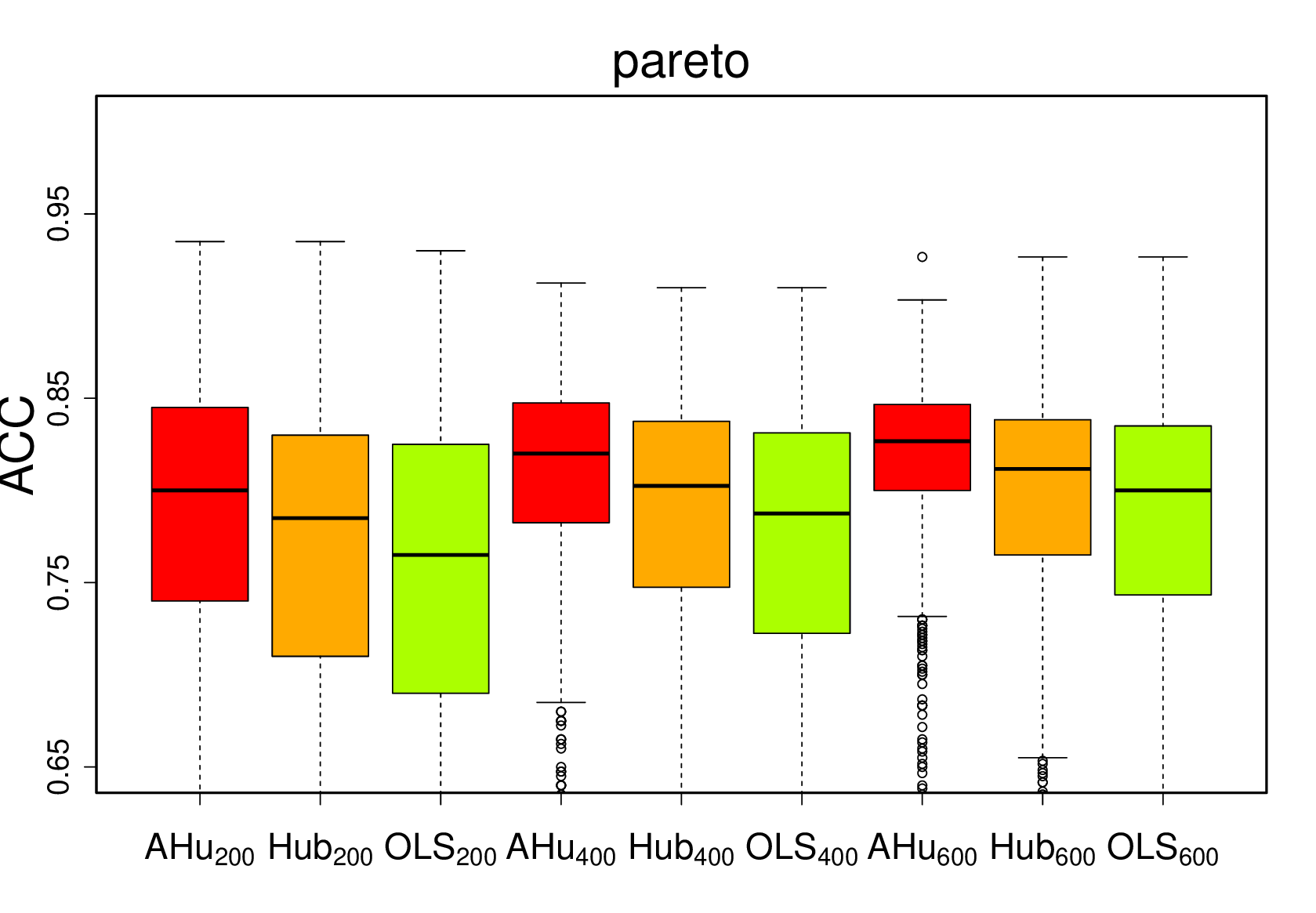}
	\includegraphics[height=2.7in,width=1.5in, angle=-90]{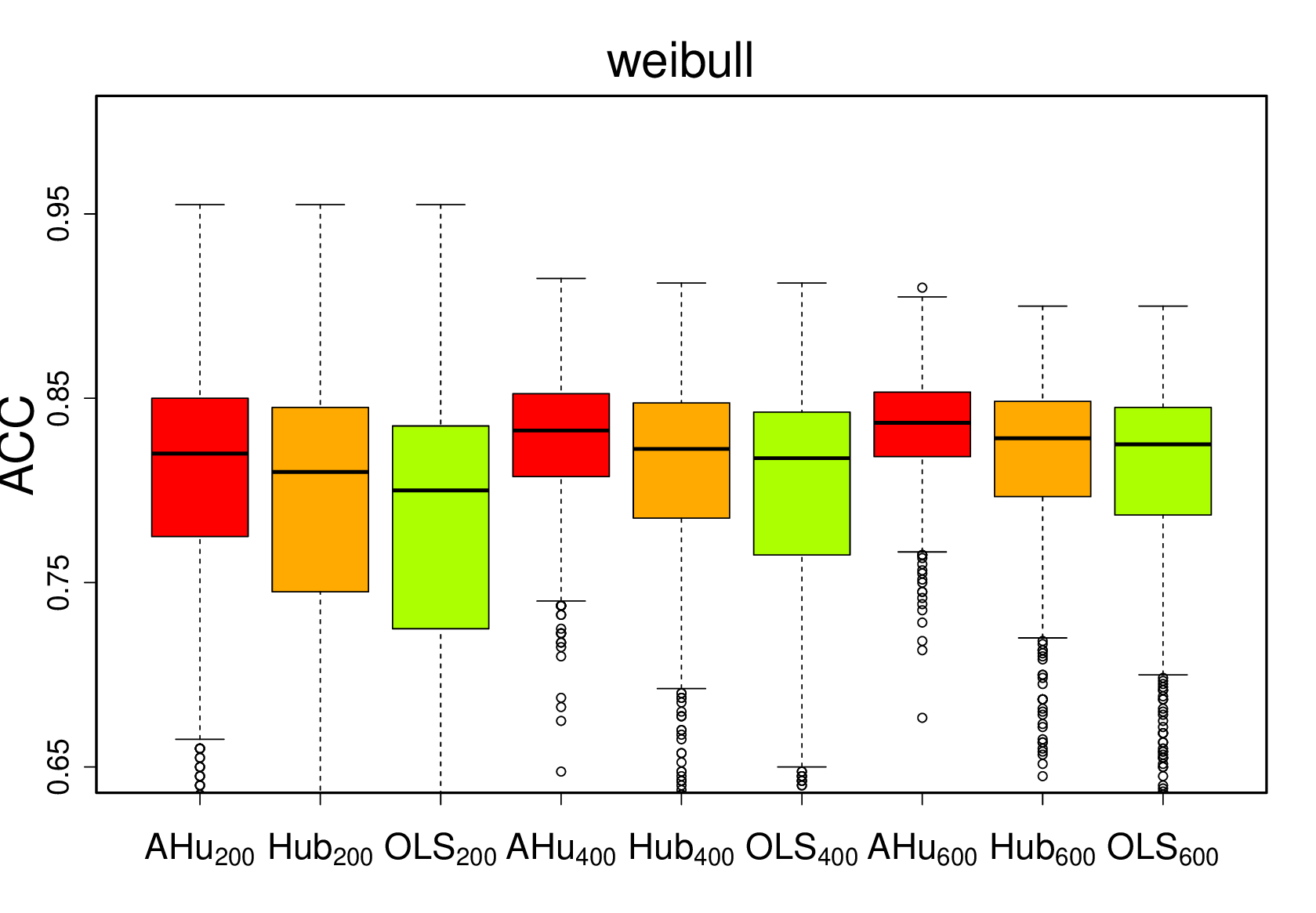}
	\caption{\it Accuracy of subgroup identification with four different error distributions. Here, $(p, q, r)=(3, 3, 3)$.}
	\label{fig_acc_qr33}
\end{center}
\end{figure}

\autoref{fig_L2_qr33} shows boxplots of the $L_2$-norm of the estimation errors of the parameter $\btheta = (\balpha\trans, \bbeta\trans)\trans$ with different errors and with $(p, q, r)=(3, 3, 3)$, where the $L_2$-norm is defined as $L_2=\|\hbtheta_{\tau,h}-\btheta^*\|^2$, where $\hbtheta_{\tau,h}$ is the robust estimator of the true parameter $\btheta^*$.
\autoref{fig_acc_qr33} shows boxplots of the accuracy (ACC) of subgroup identification with different errors and with $(p, q, r)=(3, 3, 3)$. Here, ACC is defined as $$\mbox{ACC}=1-n^{-1}\sum_{i=1}^{n}\left|\bone(U_{1i}+\bU_{2i}\trans\hbata_{\tau,h}\geq0)-\bone(U_{1i}+\bU_{2i}\trans\bata^*\geq0)\right|$$ with the robust estimator $\hbata_{\tau,h}$ of the true parameter $\bata^*$.

Figures \ref{fig_L2_qr33} and \ref{fig_acc_qr33} show that the $L_2$-norms of the estimation errors decrease and the accuracies of subgroup identification increase as the sample size $n$ grows, which verifies the established theoretical results. Compared with the ordinary quadratic loss, the proposed estimation procedure achieves a uniformly lower median of the $L_2$-norm of the estimation errors and higher accuracy of subgroup identification for all heavy-tailed distributions except for the $t_2$ distribution. The three methods are comparable for the symmetric distributions, such as the Gaussian and $t_2$ distributions. To limit the number of pages, the boxplots of the $L_2$-norm of the estimation errors of the parameter $\btheta$ and accuracy of subgroup identification for $(p, q, r)=(3, 3, 11), (6,6,3)$, and $(6,6,11)$ are shown in Figures E2--E4 and E6--E8 in Appendix~ E.1 of the Supplementary Material.

\section{Case study}\label{sec:case}

The proposed procedure is applied to cancer data for skin cutaneous melanoma (SKCM) downloaded from the Cancer Genome Atlas \href{https://tcga-data.nci.nih.gov/}{https://tcga-data.nci.nih.gov/}. SKCM is one of the most aggressive types of cancer, and its incidence, mortality, and disease burden are increasing annually \citep{2021Cancer,2019Observed}. It is believed that {\it CREG1}, {\it TMEM201}, and {\it CCL8} are skin-cancer susceptibility genes \citep{HU2020}, and the goal is to identify the subgroup of Breslow's thickness based on mutations of those sensitive genes.

Consideration is given to the three environmental factors of (i) gender, (ii) age at diagnosis, and (iii) Clark level at diagnosis (CLAD), all of which have been studied extensively in the literature. Studies such as those by \citet{Dickson2011} have found that these three environmental factors have positive effects. After removing missing values, there are 253 subjects, and the SKCM data are modeled by applying the change-plane model \eqref{model}
\begin{align*}
Y_i = \bX_i\trans\balpha + \bX_i\trans\bbeta\bone(\bU_i\trans\bgamma\geq0)+\eps_i, \quad i=1,\cdots,253,
\end{align*}
where $\bX_i = (1, \mbox{AGE}_i, \mbox{GENDER}_i, \mbox{CLAD}_i)\trans$, and
$\bU_i = (1, \mbox{CREG1}_i, \mbox{TMEM201}_i, \mbox{CCL8}_i)\trans$.
The three important genes {\it CREG1}, {\it TMEM201}, and {\it CCL8} are high correlated with breast cancer \citep{HU2020}.

Based on $B=5000$ bootstrap samples and the Huber loss, the $\pvalue$ with the proposed RWAST is $0.002$, which implies that based on the proposed method, there is a strong evidence for rejecting the null hypothesis. However, based on the ordinary quadratic loss, the $\pvalue$s with WAST and SST are 0.6454 and 0.0286, respectively, where $B=5000$ and $K=2000$. Therefore, there is no evidence for rejecting the null hypothesis for WAST and SST based on the ordinary quadratic loss.

The parametric estimators are listed in \autoref{table_SKCM}, from which the indicator function $\bone(\bU_i\trans\hbgamma\geq0)$ partitions the population into two subgroups with 121 and 132 subjects based on the Huber loss, and two subgroups with 91 and 162 subjects based on the ordinary quadratic loss. Therefore, there would appear to be a subgroup with a higher chance of skin cancer based on mutations of these three genes {\it CREG1}, {\it TMEM201}, and {\it CCL8}.

\begin{table}
\caption{\label{table_SKCM} Estimates of parameter if null hypothesis has been rejected.}
\resizebox{\textwidth}{!}{
	\begin{threeparttable}
		\begin{tabular}{cccccccccccc}\\
			\hline
			Change plane& Parameter&  Intercept& AGE& GENDER& CLAD&& Intercept & CREG1& TMEM201 & CCL8\\
			\hline
			OLS & $\hbalpha$&        $-0.037$ & $-0.038$ & 0.002 & 0.050 && && \\
			& $\hbbeta$&        0.022 & 0.024 & 0.008 & $-0.007$ && && \\
			& $\hbgamma$&          &&& && 1.000 & 0.179 & $-0.656$ & 4.970 \\
			[1 ex]
			Huber & $\hbalpha$&   3.525 & 0.029 & 0.079 & 0.064 && && \\
			& $\hbbeta$&         0.012 & 0.014 & $-0.026$ & $-0.045$ && && \\
			& $\hbgamma$&         &&& && 1.000 & 0.384 & $-2.093$ & 4.860 \\
			\hline
		\end{tabular}
	\end{threeparttable}
}
\end{table}

\section{Conclusion}\label{sec:conclus}

Considered herein were subgroup classification and subgroup tests for change-plane models with heavy-tailed errors, which offer help in (i) narrowing down populations for modeling and (ii) providing recommendations for optimal individualized treatments in practice. A novel subgroup classifier was proposed by smoothing the indicator function and minimizing a smoothed Huber loss. Nonasymptotic properties were derived and nonasymptotic Bahadur representation was provided, in which the estimators of the parameters $\balpha$ and $\bbeta$ achieve sub-Gaussian tails.

The novel test statistic RWAST was introduced to test whether subgroups of individuals exist. In a comparison with WAST \cite{Liu2022a} and SST \cite{1994Optimal,1995Admissibility, 2009On, 2017Change} based on the ordinary change-plane regression with non-heavy-tailed errors, the proposed test statistic improved the power significantly because it is robust with the assistance of the Huber loss and avoids the drawbacks of taking the supremum over the parametric space $\Theta_\gamma$ when its dimension is large. Asymptotic distributions were derived under the null and alternative hypotheses. As studied by \citet{Liu2022a} and \citet{2020Threshold}, it is easy to extend the proposed robust estimation procedure and RWAST to change-plane regression with multiple change planes with heavy-tailed errors.


In the age of big data, it is interesting to consider high-dimensional change-plane regression models and to provide high-dimensional robust estimation procedures and test statistics. As noted by \citet{Liu2022a}, the proposed RWAST can be applied to change-plane regression with high-dimensional grouping parameter $\bgamma$. However, it remains open to provide estimation procedures for change-plane regression with high-dimensional $\bgamma$. A possible strategy is to penalize the loss function with $\|\btheta\|_1$ under the assumption of sparsity. 

\section*{Supplementary Material}
Appendix~A includes Algorithm 1 for implementation in Section \ref{sec:implement}. 
Appendix~B provides the computation of critical value in Section \ref{sec:test}.
Appendix~C provides the proofs of Theorems \ref{thm1}--\ref{thm4} and the related Lemmas. 
Appendix~D provides the proofs of Theorems \ref{thm11}--\ref{thm12}. 
Appendix~E provides additional simulation studies to illustrate the performance of the proposed estimation and test prcedures.

\if0\blind{
\section*{Acknowledgments}
This work was supported in part by the NSFC (12271329, 72331005).
}\fi


\bibliographystyle{apalike}
\bibliography{papers}
\end{document}